%
%
%



\catcode `\@=11 

\def\@version{1.3}
\def\@verdate{28.11.1992}


%
%
%
%
%
%

\font\fiverm=cmr5
\font\fivei=cmmi5	\skewchar\fivei='177
\font\fivesy=cmsy5	\skewchar\fivesy='60
\font\fivebf=cmbx5

\font\sevenrm=cmr7
\font\seveni=cmmi7	\skewchar\seveni='177
\font\sevensy=cmsy7	\skewchar\sevensy='60
\font\sevenbf=cmbx7

\font\eightrm=cmr8
\font\eightbf=cmbx8
\font\eightit=cmti8
\font\eighti=cmmi8			\skewchar\eighti='177
\font\eightmib=cmmib10 at 8pt	\skewchar\eightmib='177
\font\eightsy=cmsy8			\skewchar\eightsy='60
\font\eightsyb=cmbsy10 at 8pt	\skewchar\eightsyb='60
\font\eightsl=cmsl8
\font\eighttt=cmtt8			\hyphenchar\eighttt=-1
\font\eightcsc=cmcsc10 at 8pt
\font\eightsf=cmss8

\font\ninerm=cmr9
\font\ninebf=cmbx9
\font\nineit=cmti9
\font\ninei=cmmi9			\skewchar\ninei='177
\font\ninemib=cmmib10 at 9pt	\skewchar\ninemib='177
\font\ninesy=cmsy9			\skewchar\ninesy='60
\font\ninesyb=cmbsy10 at 9pt	\skewchar\ninesyb='60
\font\ninesl=cmsl9
\font\ninett=cmtt9			\hyphenchar\ninett=-1
\font\ninecsc=cmcsc10 at 9pt
\font\ninesf=cmss9

\font\tenrm=cmr10
\font\tenbf=cmbx10
\font\tenit=cmti10
\font\teni=cmmi10		\skewchar\teni='177
\font\tenmib=cmmib10	\skewchar\tenmib='177
\font\tensy=cmsy10		\skewchar\tensy='60
\font\tensyb=cmbsy10	\skewchar\tensyb='60
\font\tenex=cmex10
\font\tensl=cmsl10
\font\tentt=cmtt10		\hyphenchar\tentt=-1
\font\tencsc=cmcsc10
\font\tensf=cmss10

\font\elevenrm=cmr10 scaled \magstephalf
\font\elevenbf=cmbx10 scaled \magstephalf
\font\elevenit=cmti10 scaled \magstephalf
\font\eleveni=cmmi10 scaled \magstephalf	\skewchar\eleveni='177
\font\elevenmib=cmmib10 scaled \magstephalf	\skewchar\elevenmib='177
\font\elevensy=cmsy10 scaled \magstephalf	\skewchar\elevensy='60
\font\elevensyb=cmbsy10 scaled \magstephalf	\skewchar\elevensyb='60
\font\elevensl=cmsl10 scaled \magstephalf
\font\eleventt=cmtt10 scaled \magstephalf	\hyphenchar\eleventt=-1
\font\elevencsc=cmcsc10 scaled \magstephalf
\font\elevensf=cmss10 scaled \magstephalf

\font\fourteenrm=cmr10 scaled \magstep2
\font\fourteenbf=cmbx10 scaled \magstep2
\font\fourteenit=cmti10 scaled \magstep2
\font\fourteeni=cmmi10 scaled \magstep2		\skewchar\fourteeni='177
\font\fourteenmib=cmmib10 scaled \magstep2	\skewchar\fourteenmib='177
\font\fourteensy=cmsy10 scaled \magstep2	\skewchar\fourteensy='60
\font\fourteensyb=cmbsy10 scaled \magstep2	\skewchar\fourteensyb='60
\font\fourteensl=cmsl10 scaled \magstep2
\font\fourteentt=cmtt10 scaled \magstep2	\hyphenchar\fourteentt=-1
\font\fourteencsc=cmcsc10 scaled \magstep2
\font\fourteensf=cmss10 scaled \magstep2

\font\seventeenrm=cmr10 scaled \magstep3
\font\seventeenbf=cmbx10 scaled \magstep3
\font\seventeenit=cmti10 scaled \magstep3
\font\seventeeni=cmmi10 scaled \magstep3	\skewchar\seventeeni='177
\font\seventeenmib=cmmib10 scaled \magstep3	\skewchar\seventeenmib='177
\font\seventeensy=cmsy10 scaled \magstep3	\skewchar\seventeensy='60
\font\seventeensyb=cmbsy10 scaled \magstep3	\skewchar\seventeensyb='60
\font\seventeensl=cmsl10 scaled \magstep3
\font\seventeentt=cmtt10 scaled \magstep3	\hyphenchar\seventeentt=-1
\font\seventeencsc=cmcsc10 scaled \magstep3
\font\seventeensf=cmss10 scaled \magstep3

\def\@typeface{Computer Modern} 

\def\hexnumber@#1{\ifnum#1<10 \number#1\else
 \ifnum#1=10 A\else\ifnum#1=11 B\else\ifnum#1=12 C\else
 \ifnum#1=13 D\else\ifnum#1=14 E\else\ifnum#1=15 F\fi\fi\fi\fi\fi\fi\fi}

\def\mib{\hexnumber@\mibfam}
\def\syb{\hexnumber@\sybfam}

\def\makestrut{%
  \setbox\strutbox=\hbox{%
    \vrule height.7\baselineskip depth.3\baselineskip width 0pt}%
}

\def\bls#1{%
  \normalbaselineskip=#1%
  \normalbaselines%
  \makestrut%
}

%

\newfam\mibfam 
\newfam\sybfam 
\newfam\scfam  
\newfam\sffam  

\def\cal{\fam2}
\def\em{\ifdim\fontdimen1\font>0 \rm\else\it\fi}

\textfont3=\tenex
\scriptfont3=\tenex
\scriptscriptfont3=\tenex

\def\eightpoint{
  \def\rm{\fam0\eightrm}%
  \textfont0=\eightrm \scriptfont0=\sevenrm \scriptscriptfont0=\fiverm%
  \textfont1=\eighti  \scriptfont1=\seveni  \scriptscriptfont1=\fivei%
  \textfont2=\eightsy \scriptfont2=\sevensy \scriptscriptfont2=\fivesy%
  \textfont\itfam=\eightit\def\it{\fam\itfam\eightit}%
  \textfont\bffam=\eightbf%
    \scriptfont\bffam=\sevenbf%
      \scriptscriptfont\bffam=\fivebf%
  \def\bf{\fam\bffam\eightbf}%
  \textfont\slfam=\eightsl\def\sl{\fam\slfam\eightsl}%
  \textfont\ttfam=\eighttt\def\tt{\fam\ttfam\eighttt}%
  \textfont\scfam=\eightcsc\def\sc{\fam\scfam\eightcsc}%
  \textfont\sffam=\eightsf\def\sf{\fam\sffam\eightsf}%
  \textfont\mibfam=\eightmib%
  \textfont\sybfam=\eightsyb%
  \bls{10pt}%
}

\def\ninepoint{
  \def\rm{\fam0\ninerm}%
  \textfont0=\ninerm \scriptfont0=\sevenrm \scriptscriptfont0=\fiverm%
  \textfont1=\ninei  \scriptfont1=\seveni  \scriptscriptfont1=\fivei%
  \textfont2=\ninesy \scriptfont2=\sevensy \scriptscriptfont2=\fivesy%
  \textfont\itfam=\nineit\def\it{\fam\itfam\nineit}%
  \textfont\bffam=\ninebf%
    \scriptfont\bffam=\sevenbf%
      \scriptscriptfont\bffam=\fivebf%
  \def\bf{\fam\bffam\ninebf}%
  \textfont\slfam=\ninesl\def\sl{\fam\slfam\ninesl}%
  \textfont\ttfam=\ninett\def\tt{\fam\ttfam\ninett}%
  \textfont\scfam=\ninecsc\def\sc{\fam\scfam\ninecsc}%
  \textfont\sffam=\ninesf\def\sf{\fam\sffam\ninesf}%
  \textfont\mibfam=\ninemib%
  \textfont\sybfam=\ninesyb%
  \bls{12pt}%
}

\def\tenpoint{
  \def\rm{\fam0\tenrm}%
  \textfont0=\tenrm \scriptfont0=\sevenrm \scriptscriptfont0=\fiverm%
  \textfont1=\teni  \scriptfont1=\seveni  \scriptscriptfont1=\fivei%
  \textfont2=\tensy \scriptfont2=\sevensy \scriptscriptfont2=\fivesy%
  \textfont\itfam=\tenit\def\it{\fam\itfam\tenit}%
  \textfont\bffam=\tenbf%
    \scriptfont\bffam=\sevenbf%
      \scriptscriptfont\bffam=\fivebf%
  \def\bf{\fam\bffam\tenbf}%
  \textfont\slfam=\tensl\def\sl{\fam\slfam\tensl}%
  \textfont\ttfam=\tentt\def\tt{\fam\ttfam\tentt}%
  \textfont\scfam=\tencsc\def\sc{\fam\scfam\tencsc}%
  \textfont\sffam=\tensf\def\sf{\fam\sffam\tensf}%
  \textfont\mibfam=\tenmib%
  \textfont\sybfam=\tensyb%
  \bls{12pt}%
}

\def\elevenpoint{
  \def\rm{\fam0\elevenrm}%
  \textfont0=\elevenrm \scriptfont0=\eightrm \scriptscriptfont0=\fiverm%
  \textfont1=\eleveni  \scriptfont1=\eighti  \scriptscriptfont1=\fivei%
  \textfont2=\elevensy \scriptfont2=\eightsy \scriptscriptfont2=\fivesy%
  \textfont\itfam=\elevenit\def\it{\fam\itfam\elevenit}%
  \textfont\bffam=\elevenbf%
    \scriptfont\bffam=\eightbf%
      \scriptscriptfont\bffam=\fivebf%
  \def\bf{\fam\bffam\elevenbf}%
  \textfont\slfam=\elevensl\def\sl{\fam\slfam\elevensl}%
  \textfont\ttfam=\eleventt\def\tt{\fam\ttfam\eleventt}%
  \textfont\scfam=\elevencsc\def\sc{\fam\scfam\elevencsc}%
  \textfont\sffam=\elevensf\def\sf{\fam\sffam\elevensf}%
  \textfont\mibfam=\elevenmib%
  \textfont\sybfam=\elevensyb%
  \bls{13pt}%
}

\def\fourteenpoint{
  \def\rm{\fam0\fourteenrm}%
  \textfont0\fourteenrm  \scriptfont0\tenrm  \scriptscriptfont0\sevenrm%
  \textfont1\fourteeni   \scriptfont1\teni   \scriptscriptfont1\seveni%
  \textfont2\fourteensy  \scriptfont2\tensy  \scriptscriptfont2\sevensy%
  \textfont\itfam=\fourteenit\def\it{\fam\itfam\fourteenit}%
  \textfont\bffam=\fourteenbf%
    \scriptfont\bffam=\tenbf%
      \scriptscriptfont\bffam=\sevenbf%
  \def\bf{\fam\bffam\fourteenbf}%
  \textfont\slfam=\fourteensl\def\sl{\fam\slfam\fourteensl}%
  \textfont\ttfam=\fourteentt\def\tt{\fam\ttfam\fourteentt}%
  \textfont\scfam=\fourteencsc\def\sc{\fam\scfam\fourteencsc}%
  \textfont\sffam=\fourteensf\def\sf{\fam\sffam\fourteensf}%
  \textfont\mibfam=\fourteenmib%
  \textfont\sybfam=\fourteensyb%
  \bls{17pt}%
}

\def\seventeenpoint{
  \def\rm{\fam0\seventeenrm}%
  \textfont0\seventeenrm  \scriptfont0\elevenrm  \scriptscriptfont0\ninerm%
  \textfont1\seventeeni   \scriptfont1\eleveni   \scriptscriptfont1\ninei%
  \textfont2\seventeensy  \scriptfont2\elevensy  \scriptscriptfont2\ninesy%
  \textfont\itfam=\seventeenit\def\it{\fam\itfam\seventeenit}%
  \textfont\bffam=\seventeenbf%
    \scriptfont\bffam=\elevenbf%
      \scriptscriptfont\bffam=\ninebf%
  \def\bf{\fam\bffam\seventeenbf}%
  \textfont\slfam=\seventeensl\def\sl{\fam\slfam\seventeensl}%
  \textfont\ttfam=\seventeentt\def\tt{\fam\ttfam\seventeentt}%
  \textfont\scfam=\seventeencsc\def\sc{\fam\scfam\seventeencsc}%
  \textfont\sffam=\seventeensf\def\sf{\fam\sffam\seventeensf}%
  \textfont\mibfam=\seventeenmib%
  \textfont\sybfam=\seventeensyb%
  \bls{20pt}%
}

\lineskip=1pt      \normallineskip=\lineskip
\lineskiplimit=0pt \normallineskiplimit=\lineskiplimit




\def\Nulle{0}  
\def\Aue{1}    
\def\Afe{2}    
\def\Ace{3}    
\def\Sue{4}    
\def\Hae{5}    
\def\Hbe{6}    
\def\Hce{7}    
\def\Hde{8}    
\def\Kwe{9}    
\def\Txe{10}   
\def\Lie{11}   
\def\Bbe{12}   


\newdimen\DimenA
\newbox\BoxA

\newcount\LastMac \LastMac=\Nulle
\newcount\HeaderNumber \HeaderNumber=0
\newcount\DefaultHeader \DefaultHeader=\HeaderNumber
\newskip\Indent

\newskip\half      \half=5.5pt plus 1.5pt minus 2.25pt
\newskip\one       \one=11pt plus 3pt minus 5.5pt
\newskip\onehalf   \onehalf=16.5pt plus 5.5pt minus 8.25pt
\newskip\two       \two=22pt plus 5.5pt minus 11pt

\def\Half{\vskip-\lastskip\vskip\half}
\def\One{\vskip-\lastskip\vskip\one}
\def\OneHalf{\vskip-\lastskip\vskip\onehalf}
\def\Two{\vskip-\lastskip\vskip\two}


\def\rTenPT{10pt plus \Feathering}

\def\TenPT{10pt plus \Feathering} 
\def\ElevenPT{11pt plus \Feathering}

\def\Raggedright{
 \rightskip=0pt plus \hsize
}

\def\Fullout{
\rightskip=0pt
}

\def\Hang#1#2{
 \hangindent=#1
 \hangafter=#2
}

\def\EveryMac{
 \Fullout
 \everypar{}
}



\def\title#1{
 \EveryMac
 \LastMac=\Nulle
 \global\HeaderNumber=0
 \global\DefaultHeader=1
 \vbox to 1pc{\vss}
 \seventeenpoint
 \Raggedright
 \noindent \bf #1
}

\def\author#1{
 \EveryMac
 \ifnum\LastMac=\Afe \OneHalf
  \else \Two
 \fi
 \LastMac=\Aue
 \fourteenpoint
 \Raggedright
 \noindent \rm #1\par
 \vskip 3pt\relax
}

\def\affiliation#1{
 \EveryMac
 \LastMac=\Afe
 \eightpoint\bls{\TenPT}
 \Raggedright
 \noindent \it #1\par
}

\def\acceptedline#1{
 \EveryMac
 \Two
 \LastMac=\Ace
 \eightpoint\bls{\TenPT}
 \Raggedright
 \noindent \rm #1
}

\def\abstract{%
 \EveryMac
 \Two
 \LastMac=\Sue
 \everypar{\Hang{11pc}{0}}
 \noindent\ninebf ABSTRACT\par
 \tenpoint\bls{\ElevenPT}
 \Fullout
 \noindent\rm
}

\def\keywords{
 \EveryMac
 \Half
 \LastMac=\Kwe
 \everypar{\Hang{11pc}{0}}
 \tenpoint\bls{\ElevenPT}
 \Fullout
 \noindent\hbox{\bf Key words:\ }
 \rm
}


\def\maketitle{%
  \Two%
  \EndOpening%
  \MakePage%
}


\def\pageoffset#1#2{\hoffset=#1\relax\voffset=#2\relax}


\def\Autonumber{
 \global\AutoNumbertrue  
}

\newif\ifAutoNumber \AutoNumberfalse
\newcount\Sec        
\newcount\SecSec
\newcount\SecSecSec

\Sec=0

\def\:{\let\@sptoken= } \:  
\def\:{\@xifnch} \expandafter\def\: {\futurelet\@tempc\@ifnch}

\def\@ifnextchar#1#2#3{%
  \let\@tempMACe #1%
  \def\@tempMACa{#2}%
  \def\@tempMACb{#3}%
  \futurelet \@tempMACc\@ifnch%
}

\def\@ifnch{%
\ifx \@tempMACc \@sptoken%
  \let\@tempMACd\@xifnch%
\else%
  \ifx \@tempMACc \@tempMACe%
    \let\@tempMACd\@tempMACa%
  \else%
    \let\@tempMACd\@tempMACb%
  \fi%
\fi%
\@tempMACd%
}

\def\@ifstar#1#2{\@ifnextchar *{\def\@tempMACa*{#1}\@tempMACa}{#2}}

\def\section{\@ifstar{\@ssection}{\@section}}

\def\@section#1{
 \EveryMac
 \Two
 \LastMac=\Hae
 \tenpoint\bls{\ElevenPT}
 \bf
 \Raggedright
 \ifAutoNumber
  \advance\Sec by 1
  \noindent\number\Sec\hskip 1pc \uppercase{#1}
  \SecSec=0
 \else
  \noindent \uppercase{#1}
 \fi
 \nobreak
}

\def\@ssection#1{
 \EveryMac
 \ifnum\LastMac=\Hae \Half
  \else \OneHalf
 \fi
 \LastMac=\Hae
 \tenpoint\bls{\ElevenPT}
 \bf
 \Raggedright
 \noindent\uppercase{#1}
}

\def\subsection#1{
 \EveryMac
 \ifnum\LastMac=\Hae \Half
  \else \OneHalf
 \fi
 \LastMac=\Hbe
 \tenpoint\bls{\ElevenPT}
 \bf
 \Raggedright
 \ifAutoNumber
  \advance\SecSec by 1
  \noindent\number\Sec.\number\SecSec
  \hskip 1pc #1
  \SecSecSec=0
 \else
  \noindent #1
 \fi
 \nobreak
}

\def\subsubsection#1{
 \EveryMac
 \ifnum\LastMac=\Hbe \Half
  \else \OneHalf
 \fi
 \LastMac=\Hce
 \ninepoint\bls{\ElevenPT}
 \it
 \Raggedright
 \ifAutoNumber
  \advance\SecSecSec by 1
  \noindent\number\Sec.\number\SecSec.\number\SecSecSec
  \hskip 1pc #1
 \else
  \noindent #1
 \fi
 \nobreak
}

\def\paragraph#1{
 \EveryMac
 \One
 \LastMac=\Hde
 \ninepoint\bls{\ElevenPT}
 \noindent \it #1
 \rm
}


\def\tx{
 \EveryMac
 \ifnum\LastMac=\Lie \Half\fi
 \ifnum\LastMac=\Hae \nobreak\Half\fi
 \ifnum\LastMac=\Hbe \nobreak\Half\fi
 \ifnum\LastMac=\Hce \nobreak\Half\fi
 \ifnum\LastMac=\Lie \else \noindent\fi
 \LastMac=\Txe
 \ninepoint\bls{\ElevenPT}
 \rm
}


\def\item{
 \par
 \EveryMac
 \ifnum\LastMac=\Lie
  \else \Half
 \fi
 \LastMac=\Lie
 \ninepoint\bls{\ElevenPT}
 \rm
}


\def\bibitem{
 \par
 \EveryMac
 \ifnum\LastMac=\Bbe
  \else \Half
 \fi
 \LastMac=\Bbe
 \Hang{1.5em}{1}
 \eightpoint\bls{\TenPT}
 \Raggedright
 \noindent \rm
}


\newtoks\CatchLine

\def\@journal{Mon.\ Not.\ R.\ Astron.\ Soc.\ }  
\def\@pubyear{1994}        
\def\@pagerange{000--000}  
\def\@volume{000}          
\def\@microfiche{}         %

\def\pubyear#1{\gdef\@pubyear{#1}\@makecatchline}
\def\pagerange#1{\gdef\@pagerange{#1}\@makecatchline}
\def\volume#1{\gdef\@volume{#1}\@makecatchline}
\def\microfiche#1{\gdef\@microfiche{and Microfiche\ #1}\@makecatchline}

\def\@makecatchline{%
  \global\CatchLine{%
    {\rm \@journal {\bf \@volume},\ \@pagerange\ (\@pubyear)\ \@microfiche}}%
}

\@makecatchline 

\newtoks\LeftHeader
\def\shortauthor#1{
 \global\LeftHeader{#1}
}

\newtoks\RightHeader
\def\shorttitle#1{
 \global\RightHeader{#1}
}

\def\PageHead{
 \EveryMac
 \ifnum\HeaderNumber=1 \Pagehead
  \else \Catchline
 \fi
}

\def\Catchline{%
 \vbox to 0pt{\vskip-22.5pt
  \hbox to \PageWidth{\vbox to8.5pt{}\noindent
  \eightpoint\the\CatchLine\hfill}\vss}
 \nointerlineskip
}

\def\Pagehead{%
 \ifodd\pageno
   \vbox to 0pt{\vskip-22.5pt
   \hbox to \PageWidth{\vbox to8.5pt{}\elevenpoint\it\noindent
    \hfill\the\RightHeader\hskip1.5em\rm\folio}\vss}
 \else
   \vbox to 0pt{\vskip-22.5pt
   \hbox to \PageWidth{\vbox to8.5pt{}\elevenpoint\rm\noindent
   \folio\hskip1.5em\it\the\LeftHeader\hfill}\vss}
 \fi
 \nointerlineskip
}

\def\PageFoot{} 

\def\authorcomment#1{%
  \gdef\PageFoot{%
    \nointerlineskip%
    \vbox to 22pt{\vfil%
      \hbox to \PageWidth{\elevenpoint\rm\noindent \hfil #1 \hfil}}%
  }%
}

\everydisplay{\displaysetup}

\newif\ifeqno
\newif\ifleqno

\def\displaysetup#1$${%
 \displaytest#1\eqno\eqno\displaytest
}

\def\displaytest#1\eqno#2\eqno#3\displaytest{%
 \if!#3!\ldisplaytest#1\leqno\leqno\ldisplaytest
 \else\eqnotrue\leqnofalse\def\eqn{#2}\def\eq{#1}\fi
 \generaldisplay$$}

\def\ldisplaytest#1\leqno#2\leqno#3\ldisplaytest{%
 \def\eq{#1}%
 \if!#3!\eqnofalse\else\eqnotrue\leqnotrue
  \def\eqn{#2}\fi}

\def\generaldisplay{%
\ifeqno \ifleqno 
   \hbox to \hsize{\noindent
     $\displaystyle\eq$\hfil$\displaystyle\eqn$}
  \else
    \hbox to \hsize{\noindent
     $\displaystyle\eq$\hfil$\displaystyle\eqn$}
  \fi
 \else
 \hbox to \hsize{\vbox{\noindent
  $\displaystyle\eq$\hfil}}
 \fi
}

\def\@notice{%
  \par\Two%
  \bls{12pt}%
  \noindent\tenrm This paper has been produced using the Blackwell
                  Scientific Publications \TeX\ macros.%
}

\outer\def\bye{\@notice\par\vfill\supereject\end}

\everyjob{%
  \Warn{Monthly notices of the RAS journal style (\@typeface)\space
        v\@version,\space \@verdate.}\Warn{}%
}




\newif\if@debug \@debugfalse  

\def\Print#1{\if@debug\immediate\write16{#1}\else \fi}
\def\Warn#1{\immediate\write16{#1}}
\def\wlog#1{}

\newcount\Iteration 

\newif\ifFigureBoxes  
\FigureBoxestrue

\def\Single{0} \def\Double{1}                 
\def\Figure{0} \def\Table{1}                  

\def\InStack{0}  
\def\InZoneA{1}
\def\InZoneB{2}
\def\InZoneC{3}

\newcount\TEMPCOUNT 
\newdimen\TEMPDIMEN 
\newbox\TEMPBOX     
\newbox\VOIDBOX     

\newcount\LengthOfStack 
\newcount\MaxItems      
\newcount\StackPointer
\newcount\Point         
\newcount\NextFigure    
\newcount\NextTable     
\newcount\NextItem      

\newcount\StatusStack   
\newcount\NumStack      
\newcount\TypeStack     
\newcount\SpanStack     
\newcount\BoxStack      

\newcount\ItemSTATUS    
\newcount\ItemNUMBER    
\newcount\ItemTYPE      
\newcount\ItemSPAN      
\newbox\ItemBOX         
\newdimen\ItemSIZE      

\newdimen\PageHeight    
\newdimen\TextLeading   
\newdimen\Feathering    
\newcount\LinesPerPage  
\newdimen\ColumnWidth   
\newdimen\ColumnGap     
\newdimen\PageWidth     
\newdimen\BodgeHeight   
\newcount\Leading       

\newdimen\ZoneBSize  
\newdimen\TextSize   
\newbox\ZoneABOX     
\newbox\ZoneBBOX     
\newbox\ZoneCBOX     

\newif\ifFirstSingleItem
\newif\ifFirstZoneA
\newif\ifMakePageInComplete
\newif\ifMoreFigures \MoreFiguresfalse 
\newif\ifMoreTables  \MoreTablesfalse  

\newif\ifFigInZoneB 
\newif\ifFigInZoneC 
\newif\ifTabInZoneB 
\newif\ifTabInZoneC

\newif\ifZoneAFullPage

\newbox\MidBOX    
\newbox\LeftBOX
\newbox\RightBOX
\newbox\PageBOX   

\newif\ifLeftCOL  
\LeftCOLtrue

\newdimen\ZoneBAdjust

\newcount\ItemFits
\def\Yes{1}
\def\No{2}




\MaxItems=15
\NextFigure=0        
\NextTable=1

\BodgeHeight=6pt
\TextLeading=11pt    
\Leading=11
\Feathering=0pt      
\LinesPerPage=61     
\topskip=\TextLeading
\ColumnWidth=20pc    
\ColumnGap=2pc       

\def\ItemSep{\vskip \TextLeading plus \TextLeading minus 4pt}

\FigureBoxesfalse 

\parskip=0pt
\parindent=18pt
\widowpenalty=0
\clubpenalty=10000
\tolerance=1500
\hbadness=1500
\abovedisplayskip=6pt plus 2pt minus 2pt
\belowdisplayskip=6pt plus 2pt minus 2pt
\abovedisplayshortskip=6pt plus 2pt minus 2pt
\belowdisplayshortskip=6pt plus 2pt minus 2pt

\PageHeight=\TextLeading 
\multiply\PageHeight by \LinesPerPage
\advance\PageHeight by \topskip

\PageWidth=2\ColumnWidth
\advance\PageWidth by \ColumnGap




\newcount\DUMMY \StatusStack=\allocationnumber
\newcount\DUMMY \newcount\DUMMY \newcount\DUMMY 
\newcount\DUMMY \newcount\DUMMY \newcount\DUMMY 
\newcount\DUMMY \newcount\DUMMY \newcount\DUMMY
\newcount\DUMMY \newcount\DUMMY \newcount\DUMMY 
\newcount\DUMMY \newcount\DUMMY \newcount\DUMMY

\newcount\DUMMY \NumStack=\allocationnumber
\newcount\DUMMY \newcount\DUMMY \newcount\DUMMY 
\newcount\DUMMY \newcount\DUMMY \newcount\DUMMY 
\newcount\DUMMY \newcount\DUMMY \newcount\DUMMY 
\newcount\DUMMY \newcount\DUMMY \newcount\DUMMY 
\newcount\DUMMY \newcount\DUMMY \newcount\DUMMY

\newcount\DUMMY \TypeStack=\allocationnumber
\newcount\DUMMY \newcount\DUMMY \newcount\DUMMY 
\newcount\DUMMY \newcount\DUMMY \newcount\DUMMY 
\newcount\DUMMY \newcount\DUMMY \newcount\DUMMY 
\newcount\DUMMY \newcount\DUMMY \newcount\DUMMY 
\newcount\DUMMY \newcount\DUMMY \newcount\DUMMY

\newcount\DUMMY \SpanStack=\allocationnumber
\newcount\DUMMY \newcount\DUMMY \newcount\DUMMY 
\newcount\DUMMY \newcount\DUMMY \newcount\DUMMY 
\newcount\DUMMY \newcount\DUMMY \newcount\DUMMY 
\newcount\DUMMY \newcount\DUMMY \newcount\DUMMY 
\newcount\DUMMY \newcount\DUMMY \newcount\DUMMY

\newbox\DUMMY   \BoxStack=\allocationnumber
\newbox\DUMMY   \newbox\DUMMY \newbox\DUMMY 
\newbox\DUMMY   \newbox\DUMMY \newbox\DUMMY 
\newbox\DUMMY   \newbox\DUMMY \newbox\DUMMY 
\newbox\DUMMY   \newbox\DUMMY \newbox\DUMMY 
\newbox\DUMMY   \newbox\DUMMY \newbox\DUMMY

\def\wlog{\immediate\write-1}


\def\GetItemAll#1{%
 \GetItemSTATUS{#1}
 \GetItemNUMBER{#1}
 \GetItemTYPE{#1}
 \GetItemSPAN{#1}
 \GetItemBOX{#1}
}

\def\GetItemSTATUS#1{%
 \Point=\StatusStack
 \advance\Point by #1
 \global\ItemSTATUS=\count\Point
}

\def\GetItemNUMBER#1{%
 \Point=\NumStack
 \advance\Point by #1
 \global\ItemNUMBER=\count\Point
}

\def\GetItemTYPE#1{%
 \Point=\TypeStack
 \advance\Point by #1
 \global\ItemTYPE=\count\Point
}

\def\GetItemSPAN#1{%
 \Point\SpanStack
 \advance\Point by #1
 \global\ItemSPAN=\count\Point
}

\def\GetItemBOX#1{%
 \Point=\BoxStack
 \advance\Point by #1
 \global\setbox\ItemBOX=\vbox{\copy\Point}
 \global\ItemSIZE=\ht\ItemBOX
 \global\advance\ItemSIZE by \dp\ItemBOX
 \TEMPCOUNT=\ItemSIZE
 \divide\TEMPCOUNT by \Leading
 \divide\TEMPCOUNT by 65536
 \advance\TEMPCOUNT by 1
 \ItemSIZE=\TEMPCOUNT pt
 \global\multiply\ItemSIZE by \Leading
}


\def\JoinStack{%
 \ifnum\LengthOfStack=\MaxItems 
  \Warn{WARNING: Stack is full...some items will be lost!}
 \else
  \Point=\StatusStack
  \advance\Point by \LengthOfStack
  \global\count\Point=\ItemSTATUS
  \Point=\NumStack
  \advance\Point by \LengthOfStack
  \global\count\Point=\ItemNUMBER
  \Point=\TypeStack
  \advance\Point by \LengthOfStack
  \global\count\Point=\ItemTYPE
  \Point\SpanStack
  \advance\Point by \LengthOfStack
  \global\count\Point=\ItemSPAN
  \Point=\BoxStack
  \advance\Point by \LengthOfStack
  \global\setbox\Point=\vbox{\copy\ItemBOX}
  \global\advance\LengthOfStack by 1
  \ifnum\ItemTYPE=\Figure 
   \global\MoreFigurestrue
  \else
   \global\MoreTablestrue
  \fi
 \fi
}


\def\LeaveStack#1{%
 {\Iteration=#1
 \loop
 \ifnum\Iteration<\LengthOfStack
  \advance\Iteration by 1
  \GetItemSTATUS{\Iteration}
   \advance\Point by -1
   \global\count\Point=\ItemSTATUS
  \GetItemNUMBER{\Iteration}
   \advance\Point by -1
   \global\count\Point=\ItemNUMBER
  \GetItemTYPE{\Iteration}
   \advance\Point by -1
   \global\count\Point=\ItemTYPE
  \GetItemSPAN{\Iteration}
   \advance\Point by -1
   \global\count\Point=\ItemSPAN
  \GetItemBOX{\Iteration}
   \advance\Point by -1
   \global\setbox\Point=\vbox{\copy\ItemBOX}
 \repeat}
 \global\advance\LengthOfStack by -1
}


\newif\ifStackNotClean

\def\CleanStack{%
 \StackNotCleantrue
 {\Iteration=0
  \loop
   \ifStackNotClean
    \GetItemSTATUS{\Iteration}
    \ifnum\ItemSTATUS=\InStack
     \advance\Iteration by 1
     \else
      \LeaveStack{\Iteration}
    \fi
   \ifnum\LengthOfStack<\Iteration
    \StackNotCleanfalse
   \fi
 \repeat}
}


\def\FindItem#1#2{%
 \global\StackPointer=-1 
 {\Iteration=0
  \loop
  \ifnum\Iteration<\LengthOfStack
   \GetItemSTATUS{\Iteration}
   \ifnum\ItemSTATUS=\InStack
    \GetItemTYPE{\Iteration}
    \ifnum\ItemTYPE=#1
     \GetItemNUMBER{\Iteration}
     \ifnum\ItemNUMBER=#2
      \global\StackPointer=\Iteration
      \Iteration=\LengthOfStack 
     \fi
    \fi
   \fi
  \advance\Iteration by 1
 \repeat}
}


\def\FindNext{%
 \global\StackPointer=-1 
 {\Iteration=0
  \loop
  \ifnum\Iteration<\LengthOfStack
   \GetItemSTATUS{\Iteration}
   \ifnum\ItemSTATUS=\InStack
    \GetItemTYPE{\Iteration}
   \ifnum\ItemTYPE=\Figure
    \ifMoreFigures
      \global\NextItem=\Figure
      \global\StackPointer=\Iteration
      \Iteration=\LengthOfStack 
    \fi
   \fi
   \ifnum\ItemTYPE=\Table
    \ifMoreTables
      \global\NextItem=\Table
      \global\StackPointer=\Iteration
      \Iteration=\LengthOfStack 
    \fi
   \fi
  \fi
  \advance\Iteration by 1
 \repeat}
}


\def\ChangeStatus#1#2{%
 \Point=\StatusStack
 \advance\Point by #1
 \global\count\Point=#2
}



\def\Zone{\InZoneA}

\ZoneBAdjust=0pt

\def\MakePage{
 \global\ZoneBSize=\PageHeight
 \global\TextSize=\ZoneBSize
 \global\ZoneAFullPagefalse
 \global\topskip=\TextLeading
 \MakePageInCompletetrue
 \MoreFigurestrue
 \MoreTablestrue
 \FigInZoneBfalse
 \FigInZoneCfalse
 \TabInZoneBfalse
 \TabInZoneCfalse
 \global\FirstSingleItemtrue
 \global\FirstZoneAtrue
 \global\setbox\ZoneABOX=\box\VOIDBOX
 \global\setbox\ZoneBBOX=\box\VOIDBOX
 \global\setbox\ZoneCBOX=\box\VOIDBOX
 \loop
  \ifMakePageInComplete
 \FindNext
 \ifnum\StackPointer=-1
  \NextItem=-1
  \MoreFiguresfalse
  \MoreTablesfalse
 \fi
 \ifnum\NextItem=\Figure
   \FindItem{\Figure}{\NextFigure}
   \ifnum\StackPointer=-1 \global\MoreFiguresfalse
   \else
    \GetItemSPAN{\StackPointer}
    \ifnum\ItemSPAN=\Single \def\Zone{\InZoneB}\relax
     \ifFigInZoneC \global\MoreFiguresfalse\fi
    \else
     \def\Zone{\InZoneA}
     \ifFigInZoneB \def\Zone{\InZoneC}\fi
    \fi
   \fi
   \ifMoreFigures\Print{}\FigureItems\fi
 \fi
\ifnum\NextItem=\Table
   \FindItem{\Table}{\NextTable}
   \ifnum\StackPointer=-1 \global\MoreTablesfalse
   \else
    \GetItemSPAN{\StackPointer}
    \ifnum\ItemSPAN=\Single\relax
     \ifTabInZoneC \global\MoreTablesfalse\fi
    \else
     \def\Zone{\InZoneA}
     \ifTabInZoneB \def\Zone{\InZoneC}\fi
    \fi
   \fi
   \ifMoreTables\Print{}\TableItems\fi
 \fi
   \MakePageInCompletefalse 
   \ifMoreFigures\MakePageInCompletetrue\fi
   \ifMoreTables\MakePageInCompletetrue\fi
 \repeat
 \ifZoneAFullPage
  \global\TextSize=0pt
  \global\ZoneBSize=0pt
  \global\vsize=0pt\relax
  \global\topskip=0pt\relax
  \vbox to 0pt{\vss}
  \eject
 \else
 \global\advance\ZoneBSize by -\ZoneBAdjust
 \global\vsize=\ZoneBSize
 \global\hsize=\ColumnWidth
 \global\ZoneBAdjust=0pt
 \ifdim\TextSize<23pt
 \Warn{}
 \Warn{* Making column fall short: TextSize=\the\TextSize *}
 \vskip-\lastskip\eject\fi
 \fi
}

\def\MakeRightCol{
 \global\TextSize=\ZoneBSize
 \MakePageInCompletetrue
 \MoreFigurestrue
 \MoreTablestrue
 \global\FirstSingleItemtrue
 \global\setbox\ZoneBBOX=\box\VOIDBOX
 \def\Zone{\InZoneB}
 \loop
  \ifMakePageInComplete
 \FindNext
 \ifnum\StackPointer=-1
  \NextItem=-1
  \MoreFiguresfalse
  \MoreTablesfalse
 \fi
 \ifnum\NextItem=\Figure
   \FindItem{\Figure}{\NextFigure}
   \ifnum\StackPointer=-1 \MoreFiguresfalse
   \else
    \GetItemSPAN{\StackPointer}
    \ifnum\ItemSPAN=\Double\relax
     \MoreFiguresfalse\fi
   \fi
   \ifMoreFigures\Print{}\FigureItems\fi
 \fi
 \ifnum\NextItem=\Table
   \FindItem{\Table}{\NextTable}
   \ifnum\StackPointer=-1 \MoreTablesfalse
   \else
    \GetItemSPAN{\StackPointer}
    \ifnum\ItemSPAN=\Double\relax
     \MoreTablesfalse\fi
   \fi
   \ifMoreTables\Print{}\TableItems\fi
 \fi
   \MakePageInCompletefalse 
   \ifMoreFigures\MakePageInCompletetrue\fi
   \ifMoreTables\MakePageInCompletetrue\fi
 \repeat
 \ifZoneAFullPage
  \global\TextSize=0pt
  \global\ZoneBSize=0pt
  \global\vsize=0pt\relax
  \global\topskip=0pt\relax
  \vbox to 0pt{\vss}
  \eject
 \else
 \global\vsize=\ZoneBSize
 \global\hsize=\ColumnWidth
 \ifdim\TextSize<23pt
 \Warn{}
 \Warn{* Making column fall short: TextSize=\the\TextSize *}
 \vskip-\lastskip\eject\fi
\fi
}

\def\FigureItems{
 \Print{Considering...}
 \ShowItem{\StackPointer}
 \GetItemBOX{\StackPointer} 
 \GetItemSPAN{\StackPointer}
  \CheckFitInZone 
  \ifnum\ItemFits=\Yes
   \ifnum\ItemSPAN=\Single
     \ChangeStatus{\StackPointer}{\InZoneB} 
     \global\FigInZoneBtrue
     \ifFirstSingleItem
      \hbox{}\vskip-\BodgeHeight
     \global\advance\ItemSIZE by \TextLeading
     \fi
     \unvbox\ItemBOX\ItemSep
     \global\FirstSingleItemfalse
     \global\advance\TextSize by -\ItemSIZE
     \global\advance\TextSize by -\TextLeading
   \else
    \ifFirstZoneA
     \global\advance\ItemSIZE by \TextLeading
     \global\FirstZoneAfalse\fi
    \global\advance\TextSize by -\ItemSIZE
    \global\advance\TextSize by -\TextLeading
    \global\advance\ZoneBSize by -\ItemSIZE
    \global\advance\ZoneBSize by -\TextLeading
    \ifFigInZoneB\relax
     \else
     \ifdim\TextSize<3\TextLeading
     \global\ZoneAFullPagetrue
     \fi
    \fi
    \ChangeStatus{\StackPointer}{\Zone}
    \ifnum\Zone=\InZoneC \global\FigInZoneCtrue\fi
  \fi
   \Print{TextSize=\the\TextSize}
   \Print{ZoneBSize=\the\ZoneBSize}
  \global\advance\NextFigure by 1
   \Print{This figure has been placed.}
  \else
   \Print{No space available for this figure...holding over.}
   \Print{}
   \global\MoreFiguresfalse
  \fi
}

\def\TableItems{
 \Print{Considering...}
 \ShowItem{\StackPointer}
 \GetItemBOX{\StackPointer} 
 \GetItemSPAN{\StackPointer}
  \CheckFitInZone 
  \ifnum\ItemFits=\Yes
   \ifnum\ItemSPAN=\Single
    \ChangeStatus{\StackPointer}{\InZoneB}
     \global\TabInZoneBtrue
     \ifFirstSingleItem
      \hbox{}\vskip-\BodgeHeight
     \global\advance\ItemSIZE by \TextLeading
     \fi
     \unvbox\ItemBOX\ItemSep
     \global\FirstSingleItemfalse
     \global\advance\TextSize by -\ItemSIZE
     \global\advance\TextSize by -\TextLeading
   \else
    \ifFirstZoneA
    \global\advance\ItemSIZE by \TextLeading
    \global\FirstZoneAfalse\fi
    \global\advance\TextSize by -\ItemSIZE
    \global\advance\TextSize by -\TextLeading
    \global\advance\ZoneBSize by -\ItemSIZE
    \global\advance\ZoneBSize by -\TextLeading
    \ifFigInZoneB\relax
     \else
     \ifdim\TextSize<3\TextLeading
     \global\ZoneAFullPagetrue
     \fi
    \fi
    \ChangeStatus{\StackPointer}{\Zone}
    \ifnum\Zone=\InZoneC \global\TabInZoneCtrue\fi
   \fi
  \global\advance\NextTable by 1
   \Print{This table has been placed.}
  \else
  \Print{No space available for this table...holding over.}
   \Print{}
   \global\MoreTablesfalse
  \fi
}


\def\CheckFitInZone{%
{\advance\TextSize by -\ItemSIZE
 \advance\TextSize by -\TextLeading
 \ifFirstSingleItem
  \advance\TextSize by \TextLeading
 \fi
 \ifnum\Zone=\InZoneA\relax
  \else \advance\TextSize by -\ZoneBAdjust
 \fi
 \ifdim\TextSize<3\TextLeading \global\ItemFits=\No
 \else \global\ItemFits=\Yes\fi}
}

\def\BF#1#2{
 \ItemSTATUS=\InStack
 \ItemNUMBER=#1
 \ItemTYPE=\Figure
 \if#2S \ItemSPAN=\Single
  \else \ItemSPAN=\Double
 \fi
 \setbox\ItemBOX=\vbox{}
}

\def\BT#1#2{
 \ItemSTATUS=\InStack
 \ItemNUMBER=#1
 \ItemTYPE=\Table
 \if#2S \ItemSPAN=\Single
  \else \ItemSPAN=\Double
 \fi
 \setbox\ItemBOX=\vbox{}
}

\def\BeginOpening{%
 \hsize=\PageWidth
 \global\setbox\ItemBOX=\vbox\bgroup
}

\let\begintopmatter=\BeginOpening  

\def\EndOpening{%
 \egroup
 \ItemNUMBER=0
 \ItemTYPE=\Figure
 \ItemSPAN=\Double
 \ItemSTATUS=\InStack
 \JoinStack
}


\newbox\tmpbox

\def\FC#1#2#3#4{%
  \ItemSTATUS=\InStack
  \ItemNUMBER=#1
  \ItemTYPE=\Figure
  \if#2S
    \ItemSPAN=\Single \TEMPDIMEN=\ColumnWidth
  \else
    \ItemSPAN=\Double \TEMPDIMEN=\PageWidth
  \fi
  {\hsize=\TEMPDIMEN
   \global\setbox\ItemBOX=\vbox{%
     \ifFigureBoxes
       \B{\TEMPDIMEN}{#3}
     \else
       \vbox to #3{\vfil}%
     \fi%
     \eightpoint\rm\bls{\rTenPT}%
     \vskip 5.5pt plus 6pt%
     \setbox\tmpbox=\vbox{#4\par}%
     \ifdim\ht\tmpbox>10pt 
       \noindent #4\par%
     \else
       \hbox to \hsize{\hfil #4\hfil}%
     \fi%
   }%
  }%
  \JoinStack%
  \Print{Processing source for figure {\the\ItemNUMBER}}%
}


\def\figps#1#2#3#4{%
  \ItemSTATUS=\InStack
  \ItemNUMBER=#1
  \ItemTYPE=\Figure
  \if#2S
    \ItemSPAN=\Single \TEMPDIMEN=\ColumnWidth
  \else
    \ItemSPAN=\Double \TEMPDIMEN=\PageWidth
  \fi
  {\hsize=\TEMPDIMEN
   \global\setbox\ItemBOX=\vbox{#3
     \eightpoint\rm\bls{\rTenPT}%
     \vskip 5.5pt plus 6pt%
     \setbox\tmpbox=\vbox{#4\par}%
     \ifdim\ht\tmpbox>10pt 
       \noindent #4\par%
     \else
       \hbox to \hsize{\hfil #4\hfil}%
     \fi%
   }%
  }%
  \JoinStack%
  \Print{Processing source for figure {\the\ItemNUMBER}}%
}

\def\TH#1#2#3#4{%
 \ItemSTATUS=\InStack
 \ItemNUMBER=#1
 \ItemTYPE=\Table
 \if#2S \ItemSPAN=\Single \TEMPDIMEN=\ColumnWidth
  \else \ItemSPAN=\Double \TEMPDIMEN=\PageWidth
 \fi
 {\hsize=\TEMPDIMEN
 \eightpoint\bls{\rTenPT}\rm
 \global\setbox\ItemBOX=\vbox{\noindent#3\vskip 5.5pt plus5.5pt\noindent#4}}
 \JoinStack
 \Print{Processing source for table {\the\ItemNUMBER}}
}

\let\table=\TH  

\def\UnloadZoneA{%
\FirstZoneAtrue
 \Iteration=0
  \loop
   \ifnum\Iteration<\LengthOfStack
    \GetItemSTATUS{\Iteration}
    \ifnum\ItemSTATUS=\InZoneA
     \GetItemBOX{\Iteration}
     \ifFirstZoneA \vbox to \BodgeHeight{\vfil}%
     \FirstZoneAfalse\fi
     \unvbox\ItemBOX\ItemSep
     \LeaveStack{\Iteration}
     \else
     \advance\Iteration by 1
   \fi
 \repeat
}

\def\UnloadZoneC{%
\Iteration=0
  \loop
   \ifnum\Iteration<\LengthOfStack
    \GetItemSTATUS{\Iteration}
    \ifnum\ItemSTATUS=\InZoneC
     \GetItemBOX{\Iteration}
     \ItemSep\unvbox\ItemBOX
     \LeaveStack{\Iteration}
     \else
     \advance\Iteration by 1
   \fi
 \repeat
}


\def\ShowItem#1{
  {\GetItemAll{#1}
  \Print{\the#1:
  {TYPE=\ifnum\ItemTYPE=\Figure Figure\else Table\fi}
  {NUMBER=\the\ItemNUMBER}
  {SPAN=\ifnum\ItemSPAN=\Single Single\else Double\fi}
  {SIZE=\the\ItemSIZE}}}
}

\def\ShowStack{%
 \Print{}
 \Print{LengthOfStack = \the\LengthOfStack}
 \ifnum\LengthOfStack=0 \Print{Stack is empty}\fi
 \Iteration=0
 \loop
 \ifnum\Iteration<\LengthOfStack
  \ShowItem{\Iteration}
  \advance\Iteration by 1
 \repeat
}

\def\B#1#2{%
\hbox{\vrule\kern-0.4pt\vbox to #2{%
\hrule width #1\vfill\hrule}\kern-0.4pt\vrule}
}

\def\Ref#1{\begingroup\global\setbox\TEMPBOX=\vbox{\hsize=2in\noindent#1}\endgroup
\ht1=0pt\dp1=0pt\wd1=0pt\vadjust{\vtop to 0pt{\advance
\hsize0.5pc\kern-10pt\moveright\hsize\box\TEMPBOX\vss}}}

\def\MarkRef#1{\leavevmode\thinspace\hbox{\vrule\vtop
{\vbox{\hrule\kern1pt\hbox{\vphantom{\rm/}\thinspace{\rm#1}%
\thinspace}}\kern1pt\hrule}\vrule}\thinspace}%


\output{%
 \ifLeftCOL
  \global\setbox\LeftBOX=\vbox to \ZoneBSize{\box255\unvbox\ZoneBBOX}
  \global\LeftCOLfalse
  \MakeRightCol
 \else
  \setbox\RightBOX=\vbox to \ZoneBSize{\box255\unvbox\ZoneBBOX}
  \setbox\MidBOX=\hbox{\box\LeftBOX\hskip\ColumnGap\box\RightBOX}
  \setbox\PageBOX=\vbox to \PageHeight{%
  \UnloadZoneA\box\MidBOX\UnloadZoneC}
  \shipout\vbox{\PageHead\box\PageBOX\PageFoot}
  \global\advance\pageno by 1
  \global\HeaderNumber=\DefaultHeader
  \global\LeftCOLtrue
  \CleanStack
  \MakePage
 \fi
}


\catcode `\@=12 



\catcode`@=11
\def\note#1#2{%
  \let\@sf=\empty \ifhmode\edef\@sf{\spacefactor\the\spacefactor}\/\fi
  $\m@th{}^{#1}$%
  \insert\footins\bgroup
    \eightpoint\bls{\rTenPT}\rm
    \textindent{$\m@th{}^{#1}$}%
    \interlinepenalty\interfootnotelinepenalty
    \splittopskip\ht\strutbox 
    \splitmaxdepth\dp\strutbox \floatingpenalty\@MM
    \leftskip\z@skip \rightskip\z@skip \spaceskip\z@skip \xspaceskip\z@skip
    \footstrut #2\strut\par
  \egroup
  \@sf\relax}
 

\output{%
 \ifLeftCOL
  \global\setbox\LeftBOX=\vbox to \ZoneBSize{\box255\unvbox\ZoneBBOX
    \ifvoid\footins\else
    \vskip\skip\footins\unvbox\footins\fi}
  \global\LeftCOLfalse
  \MakeRightCol
 \else
  \setbox\RightBOX=\vbox to \ZoneBSize{\box255\unvbox\ZoneBBOX
    \ifvoid\footins\else
    \vskip\skip\footins\unvbox\footins\fi}
  \setbox\MidBOX=\hbox{\box\LeftBOX\hskip\ColumnGap\box\RightBOX}
  \setbox\PageBOX=\vbox to \PageHeight{%
  \UnloadZoneA\box\MidBOX\UnloadZoneC}
  \shipout\vbox{\PageHead\box\PageBOX\PageFoot}
  \global\advance\pageno by 1
  \global\HeaderNumber=\DefaultHeader
  \global\LeftCOLtrue
  \CleanStack
  \MakePage
 \fi
}

\input psfig		


%

\newcount\eqnumber
\eqnumber=1
\def\step#1{\global\advance#1 by 1}
\def\neweq{{\rm\the\eqnumber}\step{\eqnumber}}
\def\preveq#1){\advance\eqnumber by -#1 \the\eqnumber) \advance\eqnumber by #1}


\newcount\fignumber
\fignumber=0
\def\newfig{\global\advance\fignumber by 1}


\pageoffset{-2.5pc}{0pc}


\Autonumber  


\pagerange{000--000}
\pubyear{1997}
\volume{}

\begintopmatter  

\title{Cluster mass estimation from lens magnification}
\author{Eelco van Kampen}
\affiliation{Royal Observatory Edinburgh, Blackford Hill, Edinburgh EH9 3HJ}
\affiliation{Theoretical Astrophysics Center, Juliane Maries Vej 30, 
DK-2100 K{\o}benhavn {\O}, Denmark,\ {\tt eelco@tac.dk}}

\shortauthor{E.\ van Kampen}
\shorttitle{Mass estimation from lens magnification}


\acceptedline{Accepted ... Received ...; in original form ...}

\abstract 

The surface mass density of a cluster of galaxies, and thus its total
mass, can be estimated from its lens magnification. The magnification can
be determined from the variation in number counts of its background galaxies. 
In the weak lensing approximation the surface mass density is a linear
function of the magnification. However, most observational data is
concentrated in the central parts of clusters, so one needs to go
beyond the weak lensing approximation, and consider the lens shear
as well, which is unknown from the variation in number counts alone.
Our approach is to look for approximate relations between the lens
shear and other lens properties in this strong lensing regime.

Such relations exist for simple analytical cluster models, like the
isothermal sphere, but are not generally a good description of observed
or simulated galaxy clusters. We therefore study the lensing properties
of a catalogue of numerical cluster models in order to find the best
possible approximation for the shear which still allows straightforward
determination of the surface mass density.
We show that by using such an approximation one can fairly well reconstruct
the surface mass distribution from the magnification alone. The
approximations are tested using clean magnification maps obtained directly
from simulated clusters, and also using lensed mock background galaxy
distributions in order to estimate the intrinsic uncertainties of the method. 
We demonstrate that the mass estimated using the weak lens magnification
approximation is usually at least twice the true mass. We illustrate our
technique on existing data, and show that the resulting masses compare well
to other estimates.


\keywords cosmology: theory -- dark matter -- large-scale structure of Universe
-- gravitational lensing

\maketitle  	



\section{Introduction}

\tx A rich cluster of galaxies acts as a gravitational lens on the galaxy
distribution behind it. This simple fact can be used to derive a great
deal about both the lensing cluster as well as the background galaxy
population (eg.\ Schneider, Ehlers \& Falco 1992; Fort \& Mellier 1994;
Kaiser 1996). In this paper we deal with methods that exploit the
variation in galaxy number counts caused by the lensing cluster to
obtain properties of that cluster, for example its total mass.
Broadhurst, Taylor \& Peacock (1995, BTP from here on) have shown how
this variation in number counts of background galaxies depends on the
lens magnification, and how to best obtain the latter from the former.

In order to obtain a total mass for the lens, or a mass
distribution, one has to somehow derive the surface mass density
from the lens magnification. BTP use the weak lens approximation to
relate the two directly. However, this approximation is only valid
in the outskirts of clusters, while most of the observational data is
restricted to the central parts of clusters, where lensing is strong.
We therefore need to go beyond the weak lensing
approximation to realistically estimate the cluster surface mass
density in this strong lensing regime. This means that besides the
magnification one needs the shear distribution in order to obtain
the lens convergence, and thus the surface mass density of the lens.

There are various ways of obtaining the shear distribution from
observations, like the method devised by Kaiser \& Squires (1993)
that utilizes the shearing of the background galaxy images to
estimate the tangential shear component (but not the radial one).
This method also provides a way to estimate the surface mass density,
save a constant. The problem of this unknown constant, dubbed the
'sheet-mass degeneracy' (Gorenstein, Falco \& Shapiro 1988), prevents
absolute mass measures, although the edges of the observed field can
be used to put a lower limit on the mass by requiring it to be
positive everywhere. Furthermore, Kaiser \& Squires (1993) again
assume weak lensing.
Extensions to these methods in order to break this degeneracy and/or
consider the strong lensing regime have been devised by, amongst others,
Schneider \& Seitz (1995), Kaiser (1995), Schneider (1995)
and Bartelmann et al.\ (1996).
The reliability of these and other shear methods has
been discussed by Bartelmann (1995) and Wilson, Cole \& Frenk (1996).

However, an approximation that relates the shear field to either the
magnification or the convergence field allows one
to obtain an absolute measure for the convergence, and thus the
surface mass density, from the magnification alone. This also
provides a mass estimate that is independent from other methods. 
Observationally, measuring the shapes of galaxies is more difficult than
just counting them. Thus, a route to the cluster mass from the lens
magnification alone is a major advantage: ground-based observations are
sufficient, as imaging of the galaxies is not required.

We use a sample of numerical galaxy cluster models (van Kampen \& Katgert
1997) to find heuristic relations between lens shear and lens convergence
(or lens magnification). We also consider some relations that have an
underlying assumption about the physical state of the lens, like
isotropy. 

In order to find the maximum performance of the estimators that
correspond to such approximations, we test how well one can estimate
the surface mass density from just the magnification map, as obtained
directly from a numerical cluster model. Subsequently, we test how
well the estimators work on magnification maps that were obtained
from lensed mock background galaxy distributions. We roughly follow
the same procedures as an observer would for an observed galaxy
distribution, thus mimicking most of the problems involved in the
application of the method.

Recently, Fort, Mellier \& Dantel-Fort (1997) and Taylor et al.\ (1998)
showed that a depletion in number counts can clearly be observed.
Fort et al.\ (1997) showed this most convincingly for the cluster
CL 0024+1654. They did not try to estimate the cluster surface mass
density or the total mass, however. Taylor et al.\ (1998) did estimate
a mass for A1689, and found it to be consistent with other mass measures.

The paper is outlined as follows:
in Section 2 we sketch the path from observed number counts to estimated
cluster properties. The necessary approximations and methods are introduced
in Section 3, and tested on simulated data in Section 4. We apply the
technique to published data in the literature in Section 5.

\section{The lens magnification method}

\subsection{The thin lens approximation}

\tx We summarize the main features of the thin lens approximation, following
Schneider, Ehlers \& Falco (1992) and Bartelmann \& Weiss (1994),
paying attention to those elements that are important for this paper.
For properly renormalized mass and length scales (see Bartelmann \& Weiss
1994 for details), the lens equation becomes
$${\bf y}={\bf x}-{\bf\alpha}({\bf x}) \ , \eqno(\neweq)$$
where ${\bf x}$ and ${\bf y}$ represent the lens and source planes
respectively, and ${\bf\alpha}$ is the deflection angle, being the
gradient of the lens potential $\psi$, given by
\newcount\alphakappa \alphakappa=\eqnumber
$${\bf\alpha}({\bf x}) = {\bf\nabla}\psi({\bf x}) =
  (\kappa * {\bf K})({\bf x})\ . \eqno(\neweq)$$
Here we have introduced the lens convergence $\kappa$, being the
dimensionless surface mass density $\Sigma(\xi_0 {\bf x})$ of the lens
(where $\xi_0$ scales the dimensionless ${\bf x}$ to a dimensional quantity):
$$\kappa({\bf x}) \equiv {\Sigma(\xi_0 {\bf x})/\Sigma_{\rm cr}}\ ,
	\eqno(\neweq)$$
and the kernel 
$${\bf K}({\bf x}) \equiv {1\over\pi}{{\bf x}\over|{\bf x}|^2} \eqno(\neweq)$$
with which $\kappa$ is convolved to obtain the deflection angle.
The {\it critital} surface mass density $\Sigma_{\rm cr}$ plays an important
r\^ole in lensing theory. It is defined as
\newcount\defcritden \defcritden=\eqnumber
$$\Sigma_{\rm cr} \equiv \rho_{\rm cr}{c\over3H_0} f(z_{\rm d},z_{\rm s})\
  , \eqno(\neweq)$$
where $\rho_{\rm cr}$ is the critical density of the background universe,
and $f(z_{\rm d},z_{\rm s})$ a function of the lens redshift $z_{\rm d}$
and the source redshift $z_{\rm s}$, whose expression depends on the
geometry of the universe, i.e.\ the cosmological model (see Schneider
et al.\ 1992, ch. 5).
From the deflection angle we can calculate the lens magnification and
shear through the Jacobian ${\bf A}$ of the lens mapping:
$${\bf A}({\bf x})=\Bigl({\rm d{\bf y}\over d{\bf x}}\Bigr)
	= {\bf I} - {\rm d{\bf\alpha}({\bf x})\over d{\bf x}}\ .
	\eqno(\neweq)$$
The lens magnification $\mu$ is obtained from its determinant:
$$\mu^{-1}({\bf x}) = \det{\bf A}({\bf x}) =
	A_{11}A_{22} - A_{12}A_{21}\ . \eqno(\neweq)$$
The lens shear consists of the two trace-free components 
$$\gamma_1({\bf x}) ={1\over 2}(A_{22}-A_{11})\ \ ,\ 
  \gamma_2({\bf x}) =-A_{12} = -A_{21}\ , \eqno(\neweq)$$
with the total shear given by
\newcount\defgamma \defgamma=\eqnumber
$$\eqalign{\gamma^2=&\gamma_1^2+\gamma_2^2 
		   = {1\over 4}(A_{22}-A_{11})^2+A_{12}A_{21}\cr
		   =&{1\over 4}(A_{11}+A_{22})^2 - \mu^{-1} \cr}
							\eqno(\neweq) $$
The convergence can also be expressed in terms of ${\bf A}$ through the
Poisson equation for the lensing potential:
\newcount\defkappa \defkappa=\eqnumber
$$\kappa({\bf x}) = {1\over 2}\nabla{\bf\alpha} = 
	 1 - {1\over 2}A_{11} - {1\over 2}A_{22}\ .\eqno(\neweq)$$
We can thus relate the three main lensing properties:
\newcount\mukapgam \mukapgam=\eqnumber
$$\mu^{-1}= (1-\kappa)^2-\gamma^2\ . \eqno(\neweq)$$

\subsection{Cluster mass estimation from variations in number counts}

\tx We sketch the whole path, in three distinctive stages, from observed
number counts to estimated cluster properties, notably the total mass. 
Besides describing the method, we indicate where we need to make assumptions.

\subsubsection{Variation in number counts to magnification}

\tx The presence of a lens gives rise to a variation in the number counts
of background galaxies (BTP). For galaxies counted up to a magnitude limit
$m_{\rm lim}$, we denote this variation as $N/N_0$, where $N_0$ is the
average galaxy number count for the field. Assuming that the integral
luminosity function of these galaxies can for $m<m_{\rm lim}$ be
approximated by a power-law with slope $S$, the magnification $\mu$ can
be calculated from $N/N_0$ and a maximum-likelihood analysis of the
background redshift distribution (BTP), taking into account the 
clustering of background galaxies which confuses the lensing signal
(Taylor \& Dye 1998).
In case no redshift information is available for the background galaxies,
an estimate for the absolute value of $\mu$ is given by
\newcount\muest \muest=\eqnumber
$$|\mu_{\rm est}|=(N/N_0)^{\beta}  ,\
{\rm with}\ \beta={(2.5S-1)^{-1}}\ .\eqno(\neweq)$$
In most cases $\beta$ is negative, i.e.\ we observe a depletion in number
counts due to the presence of the cluster. Note that we can only measure
the absolute value of $\mu$, so we have to set its sign, the image
parity, by hand.

Although the feasibility of actually obtaining $\mu$ from observational data
is an interesting topic for discussion, the issue of this paper is how to
proceed from the magnification to the properties of the lensing cluster. 
We therefore assume for the remainder of this paper that one can reliably
measure the lens magnification, as it has been shown to be detectable
(Fort, Mellier \& Dantel-Fort 1997; Taylor et al.\ 1998).

\subsubsection{Magnification to convergence}

\tx As is obvious from eq.\ (\the\mukapgam), in order to obtain $\kappa$
one needs to know both the lens magnification and shear.
As the lens shear can be found from the distortion
of the shapes of the background galaxies (Kaiser \& Squires 1993),
one could, in principle, combine this with the magnification found from
the variation in number counts to calculate the convergence.
However, here we like to use the latter as an {\it independent}\ determination
of the convergence, as one can also derive $\kappa$, save a constant,
directly from the lens shear (Kaiser \& Squires 1993). Just detecting
galaxies is also easier than obtaining their shapes, can be done in
larger numbers, and from the ground.
In order to get $\kappa$ from the lens magnification alone,
we need to make assumptions about the shear field. We do this by finding
approximate relations for simulated clusters, which have known
lensing properties. Once we have found a relation between the shear
and either the magnification or the convergence, we can also go straight
from the variation in number counts to an estimate for the convergence.

\subsubsection{Convergence to projected cluster mass and other properties} 

\tx The convergence of a lens depends on the redshift of the background
galaxy being lensed, so for a distribution of background galaxies one
finds a weighted average over convergences for each of these galaxies.
In order to translate this convergence to a surface mass density we
just multiply by $\Sigma_{\rm cr}$ for an {\it effective}\ source redshift,
which depends on the redshift distribution adopted.
The total projected mass is then found by integration over the surface mass density.

Many other methods for cluster mass estimation give 3D masses.
We can derive these using a relation between 2D and 3D masses for the
model cluster catalogue of van Kampen \& Katgert (1997), which is given as
a fit to the scatter plot for all models (van Kampen 1998):
$${M_{\rm 3D}\over M_{\rm 2D}}= 0.56 \tan^{-1}
  \Bigl({R\over 0.14h^{-1}{\rm Mpc}}\Bigr)\ . \eqno(\neweq)$$
The scatter around this relation is fairly large for small $R$, due
to substructure along the line-of-sight, but less than 10 per cent
for $R>0.4h^{-1}{\rm Mpc}$ (van Kampen 1998).

\section{Finding an optimal convergence estimator}

\subsection{General strategy}

\tx The convergence $\kappa$ of a lens, from which we can obtain its mass,
is not just a function of the lens magnification $\mu$ that we measure, but
also of its lens shear $\gamma$ (see eq.\ \the\mukapgam).
One way to eliminate the dependence on $\gamma$ is to find
an approximate relation between $\gamma$ and $\kappa$.
This can be done by looking more closely at the Jacobian of the lens
mapping, ${\bf A}$.
All three main lensing properties are a function of two or more of its
components. One can therefore try to statistically relate these components
to each other. For example, is we assume that $A_{11}=A_{22}$,
and both $A_{12}$ and $A_{21}$ vanish, we have $\gamma=0$ and
$\mu^{-1}=(\kappa-1)^2$ (discussed in more detail below).

Another approach is to start from eq.\ \the\mukapgam, and assume an
arbitrary local relation $\gamma(\kappa)$, i.e.
$$\mu^{-1}(\kappa) = (1-\kappa)^2-\gamma^2(\kappa)\ . \eqno(\neweq)$$
For a typically aspherical and clumpy cluster, the convergence has
a strong dipole component, while the shear is dominated by a quadrupole
component. In other words, only specific lensing potentials will satisfy
such a relation exactly. However, such a relation can be a good approximation
for averaged quantities, like radial profiles, for example.
Also, when these functions are smoothed significantly, as is often the case
for observational data, approximate local relations should exist.

\newfig \newcount\Fsurface \Fsurface=\fignumber
\figps{\fignumber}{S}
{\psfig{file=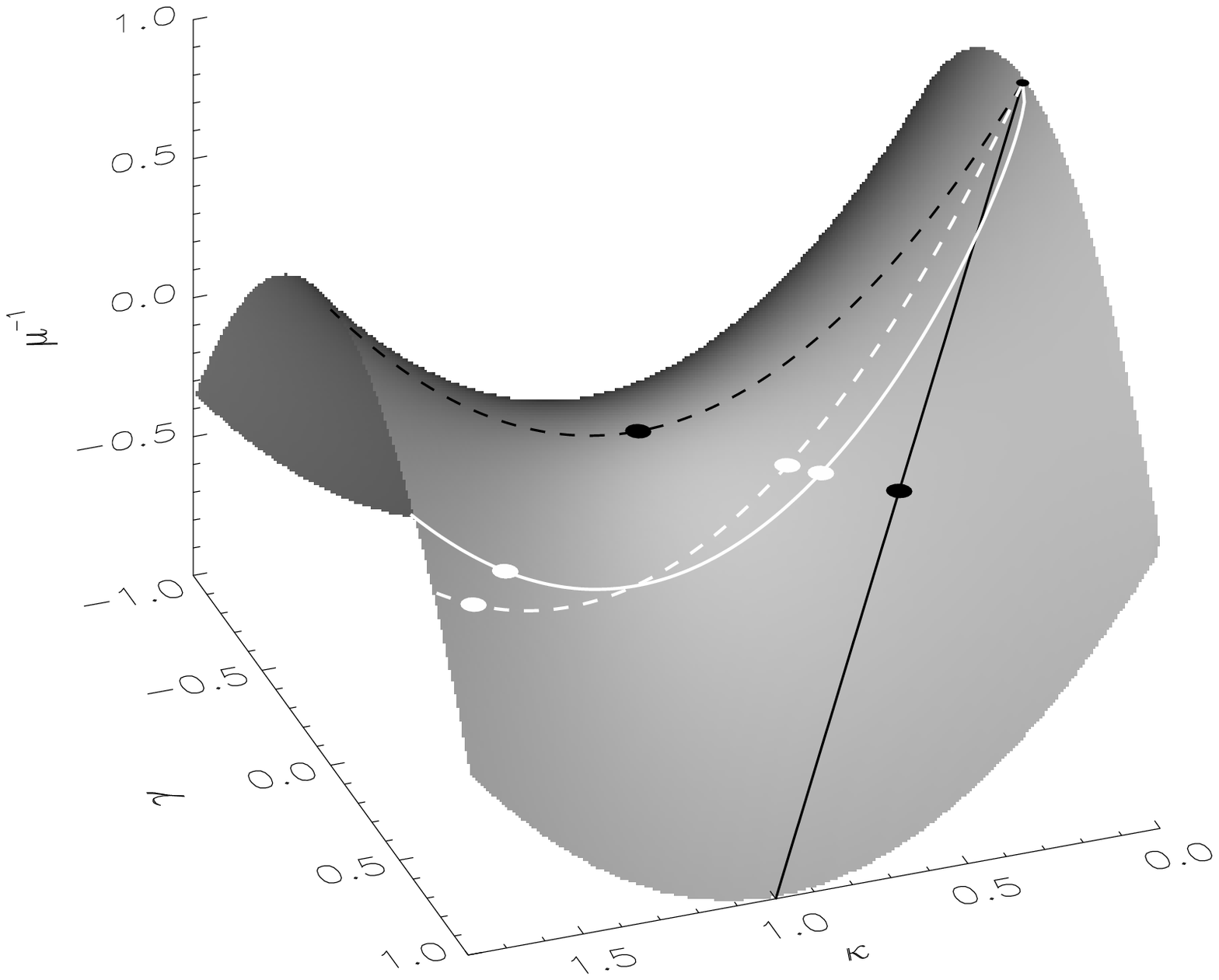,width=9cm,silent=}
}{{\bf Figure \the\fignumber.} Lens magnification $\mu^{-1}$ as a function of
convergence $\kappa$ and shear $\gamma$. Several approximations relating
$\kappa$ and $\gamma$ locally are indicated as lines on the surface,
where black lines indicate the one-parity approximations $\gamma=0$
(dashed line) and $\gamma=\kappa$ (solid line), described in
Section 3.2, and white lines the two-parity approximations
$\gamma\propto\kappa$ (dashed line) and $\gamma\propto\kappa^{1/2}$ 
(dotted line), described in Section 3.3. The caustics are indicated by
large dots.}

The assumption of locality of the shear allows many possible
specific approximations, so the aim is to select either
physically motivated $\gamma(\kappa)$, or a $\gamma(\kappa)$ that leads
to a relation between $\mu$ and $\kappa$ which is easily
invertible, and therefore applicable to observations. We use a
representative sample of cluster models to find such a relation, by
simply investigating how shear and convergence relate to each other for
these models. 

One problem we will always have to deal with is that of image parity,
the sign of the lens magnification, as we can only obtain $|\mu|$ from the
observed number counts, and have to make an educated guess about this
parity. In the general case we have a second parity as well,
as the magnification is a quadratic function of $\kappa$ and $\gamma$.
In looking for a local relation between $\kappa$ and $\gamma$ we should
therefore try to minimize the range of $\mu$ for which we have to
set parities.

In devising approximations we need to take care that the shear $\gamma$
remains real for all $\mu$. Furthermore, we set $\mu^{-1}(0)=1$, i.e.\ for 
$\kappa\rightarrow 0$ we also have $\gamma\rightarrow 0$, which
corresponds to saying that the cluster is isolated.

The various possibilities, along with their physical interpretation,
are discussed below. Both $\kappa$ and $\gamma$ are labelled to denote
the approximation made, while $\mu$ and $N/N_0$ appear unindexed, as we
consider them to be observed functions for the sake of this paper.
We first consider approximations with just one parity.

\subsection{Estimators with one parity}

\tx Estimators with only one parity which also have physical shear
distributions are special cases of eq.\ (\the\mukapgam):
the expression $(\kappa-1)^2-\gamma^2$ is allowed to have a sign
uncertainty, but its associated parity should be the same as that of
$|\mu|$ in order to effectively have one parity only.
This means that the following simple possibilities remain:
$$\eqalign{ \mu^{-1}=(\kappa-1)^2 \hskip 1.0cm & (\gamma=0) \cr
	\mu^{-1}=1-2\kappa \hskip 1.0cm & (\gamma=\kappa) \cr
	\mu^{-1}=\gamma^2  \hskip 1.0cm & (\kappa=0) \cr
  } \eqno(\neweq)$$
If we try $\mu^{-1}=(\kappa-q)^2$, then $\gamma^2=q^2-1+2(1-q)\kappa$,
which is only positive definite for $q=1$, the first possibility listed
above. The last possibility is not very likely in reality, obviously.

In Fig.\ \the\Fsurface\ we have plotted $\mu^{-1}$ as a function
of both $\kappa$ and $\gamma$. On this surface of possible
$(\kappa, \gamma, \mu^{-1})$, we have drawn the $\gamma=\kappa$
(solid black line) and $\gamma=0$ (dashed black line) approximations.
This shows why only these two remain when we require the one parity
approximations to go through $(\kappa, \gamma, \mu^{-1})=(0,0,1)$.

More complicated functions of $\mu^{-1}(\kappa)$ can be proposed, of course,
which start at $(0,0,1)$, and cross the $\mu^{-1}=0$ plane only once.
But all of these also lead to complicated (and probably multi-valued)
expressions for $\gamma(\kappa)$, and, more importantly, inversion of
$\mu^{-1}(\kappa)$ becomes less straightforward.

\subsubsection{No shear: $\gamma=0$}

\tx The first of the two one-parity approximations is also the simplest:
we forget about shear altogether, which corresponds to treating the cluster
as a uniform sheet of matter. This means that we set $A_{11}=A_{22}$, and
$A_{12}=A_{21}=0$. Setting $\gamma=0$, the estimator is easily derived
from eq.\ (\the\mukapgam):
\newcount\kappanoshear \kappanoshear=\eqnumber
$$\kappa_0 = 1 - {\cal P} |\mu|^{-1/2}\ 
	   = 1 - {\cal P} (N/N_0)^{-\beta/2}, \eqno(\neweq)$$
where ${\cal P}$ is the image parity, i.e.\ the sign of $\mu$.
In this approximation there is one critical
line, at $\kappa=1$, which separates the two parity regimes:
${\cal P}=1$ for $\kappa<1$, and ${\cal P}=-1$ for $\kappa>1$.

For observational data $\kappa$ is of course a-priori unknown, but the
position of the critical line can usually be guessed from the occurance
of giant arcs, or the position of a significant dip in the number counts,
most easily in number counts in spherical bins, but also in 2D maps.

\subsubsection{Isotropic approximation: $\gamma=\kappa$}

\tx BTP argue that if the fluctuations around the mean lensing
potential are reasonably isotropic, that :
$$\left<(1-A_{11})^2\right>
	\ \approx\ \left<(1-A_{22})^2\right>
	\ \approx\ \left<(A_{12})^2\right> \  \eqno(\neweq)$$
(note that there is a typographical error in their eq.\ 19).
If we make this exact, i.e.\
$1-A_{11}=1-A_{22}=A_{12}=A_{21}$, we find that $\gamma_1=0$, 
$\gamma_2^2=\gamma^2=\kappa^2=(1-A_{11})^2$ {\it etc.}, and
$\mu^{-1}=1-2\kappa$.

So, assuming that the shear is equal to the convergence results in the
estimator
\newcount\kappaiso \kappaiso=\eqnumber
$$\kappa_1 = {1\over 2} - {1\over 2} {\cal P} |\mu^{-1}| 
  = {1\over 2} - {1\over 2} {\cal P} \Bigl({N\over N_0}\Bigr)^{-\beta}\ 
	,\eqno(\neweq)$$
where ${\cal P}$ is again the image parity.
The critical line is now assumed to be at $\kappa=1/2$, so this estimator
automatically gives a smaller mass than the shearless mass estimator which
assumes $\kappa=1$ at the critical line. This makes physical sense, as there
is now a shear contribution to the lensing, while in the shearless case
$\kappa$ has to account for all the magnification. 

\newfig \newcount\Flensex \Flensex=\fignumber
\figps{\fignumber}{D}{\psfig{file=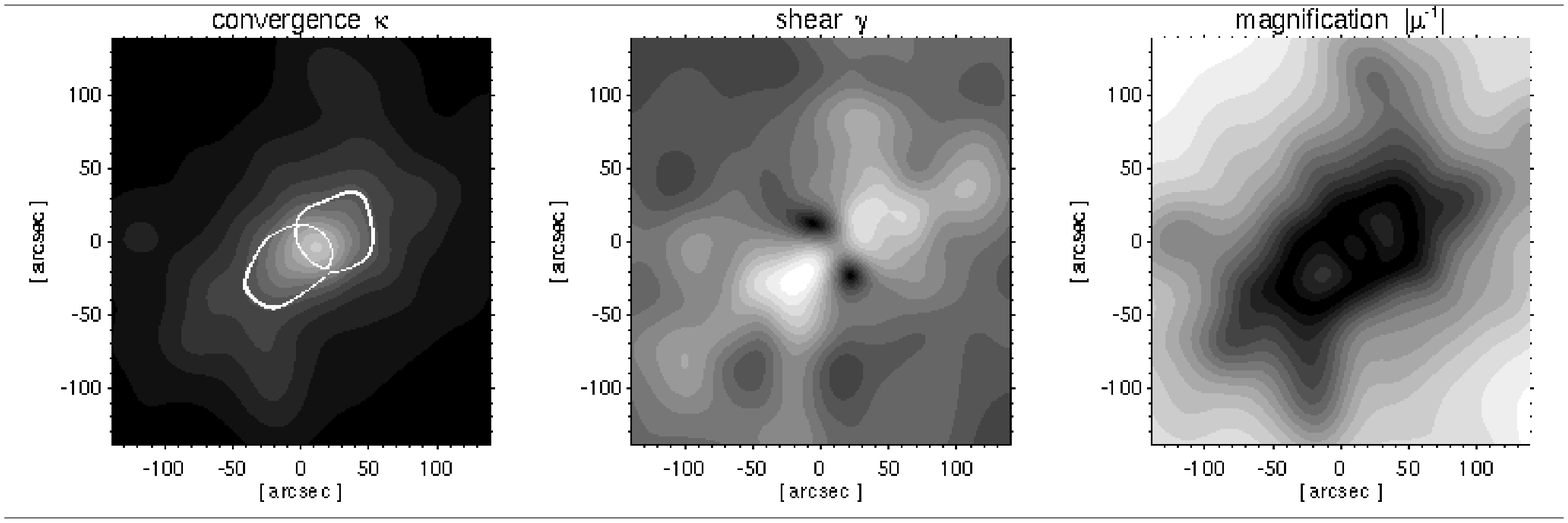,width=18cm,silent=}}
{{\bf Figure \the\fignumber.} An example of the lens convergence, shear,
and magnification for a simulated galaxy cluster (the fourth entry in
Table 1). The convergence $\kappa$ was obtained from an N-body simulation
using adaptive window smoothing (see text for details). The shear $\gamma$
and magnification $\mu$ were obtained from $\kappa$ using the thin lens
approximation.}

\subsubsection{Linear approximation}

\tx All approximations discussed so far can be linearized to the simple form
$$\kappa_{\rm lin} = {1\over 2}(\mu-1) =
  {1\over 2}\Bigl({N\over N_0}\Bigr)^{\beta}\ , \eqno(\neweq)$$
a form used by BTP and others. This approximation is obviously useful
for $\mu\approx 1$ only, corresponding to small $\kappa$, i.e.\ the outskirts
of clusters. However, most observational data is restricted to the core
of the cluster only, because of the limited field of view (for the
{\it Hubble Space Telescope}\ for example), or because the data was
taken for other reasons. This renders the linear approximation quite
useless for our purposes. We will show it for reference only in the
remainder of this paper. Note that the linear approximation has no
critical curves and no parity changes, as $\mu$, being proportional
to $\kappa$, will never become zero; it will only increase towards
the centre.

\newfig \newcount\Fkapgamex \Fkapgamex=\fignumber
\figps{\fignumber}{D}{\psfig{file=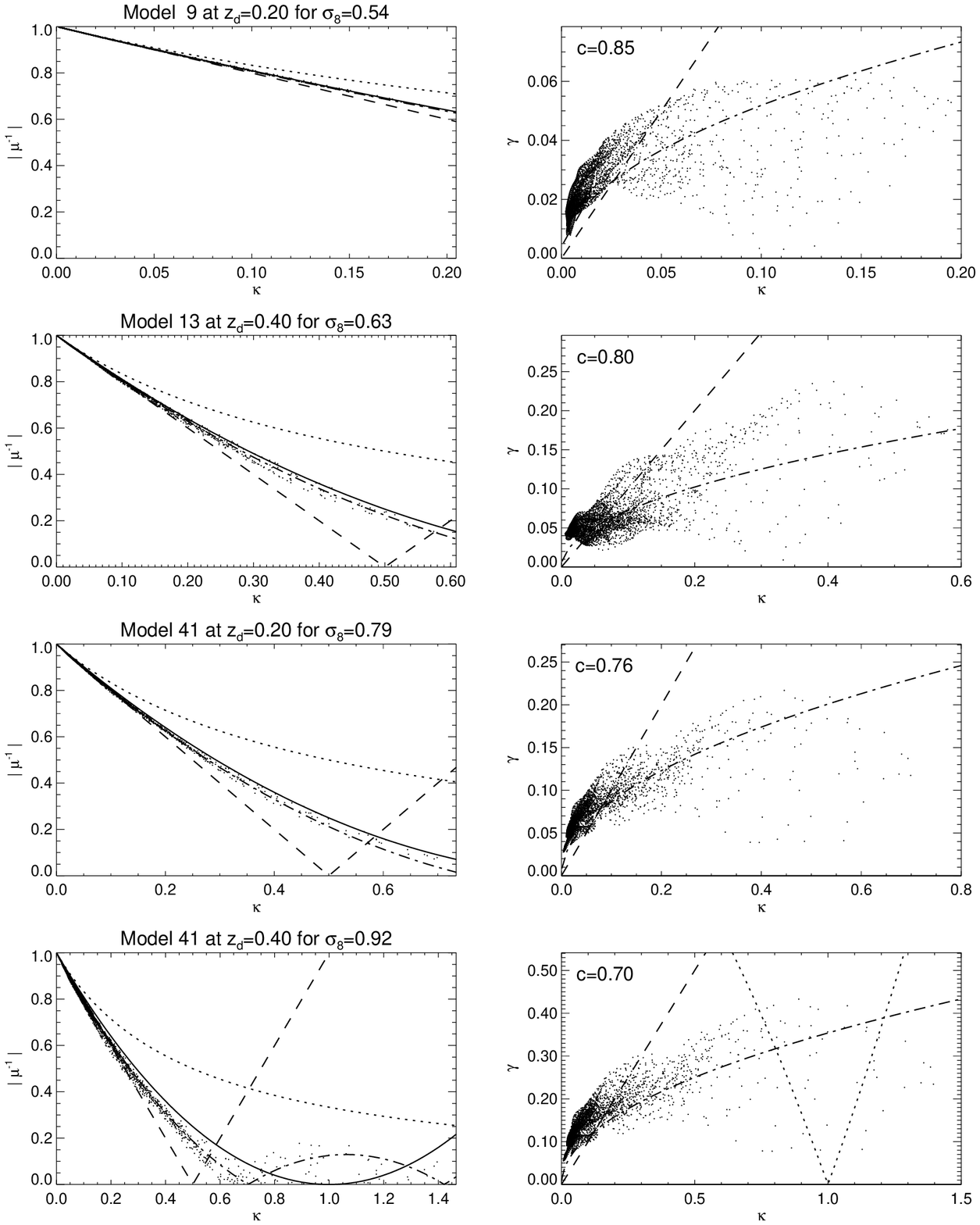,width=17.0cm,silent=}}
{{\bf Figure \the\fignumber.} The left column shows scatter plots of
the lens magnification $\mu$ versus the lens convergence $\kappa$,
with corresponding plots of the lens shear $\gamma$ versus $\kappa$, for
the four clusters listed in Table1. The parameter $c$ for the
$\gamma\propto\kappa^{1/2}$ estimator (dot$^3$-dashed line) was fitted
using the $\kappa$-$\gamma$ plots, and is annotated in the right hand
panels.
Solid lines correspond to the $\gamma=0$ approximation, dashed
lines are for $\gamma=\kappa$, and dotted lines indicate the weak
(i.e.\ linear) lens approximation.}

\newfig \newcount\Fapprox \Fapprox=\fignumber
\figps{\fignumber}{S}{\psfig{file=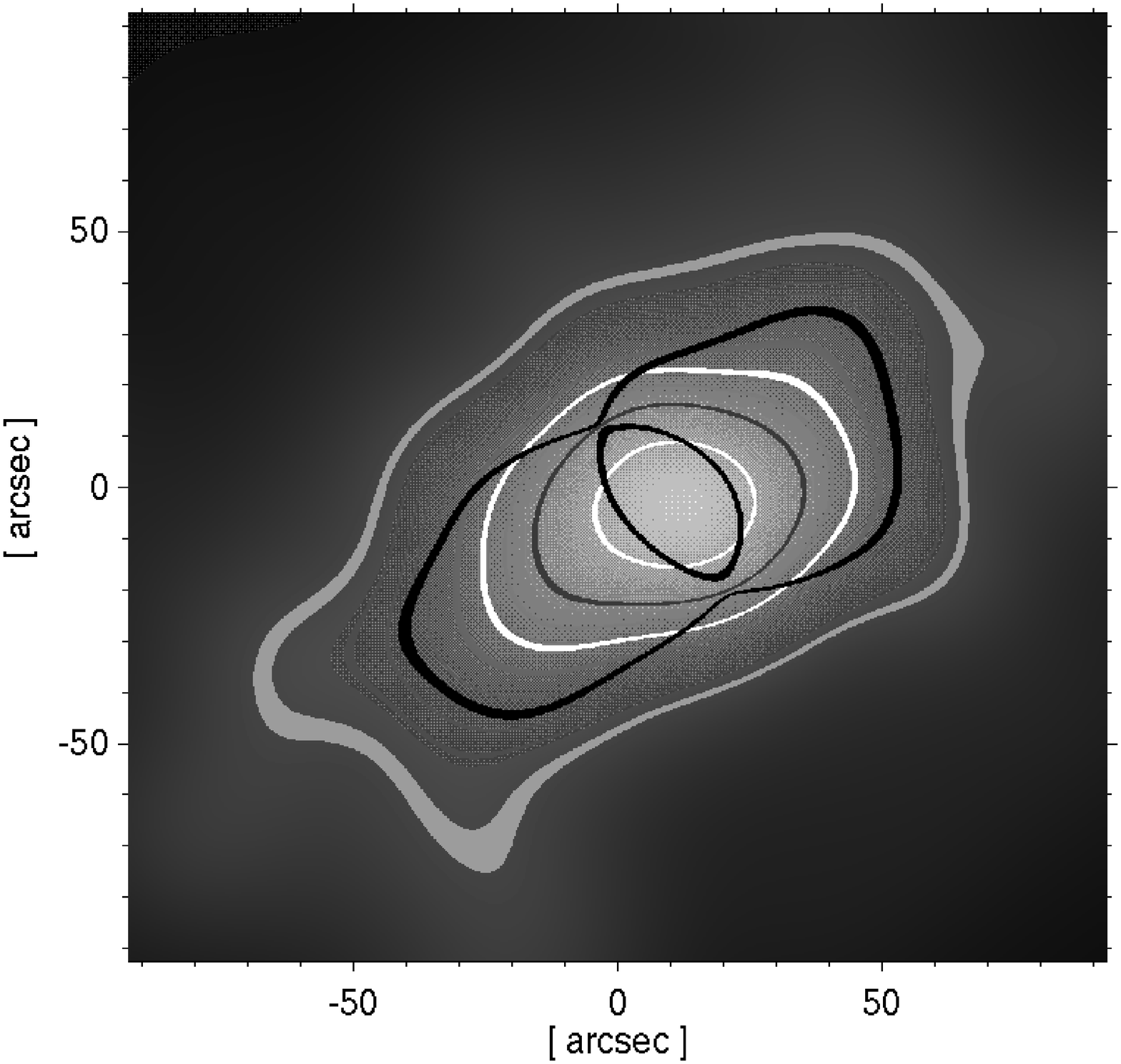,width=8.7cm,silent=}}
{{\bf Figure \the\fignumber.} Comparison of the shear approximations in
terms of their caustic lines. Show is the lens convergence for a massive
clusters (the fourth entry of Table 1), with its true caustic lines
(black curves) and the caustic lines corresponding to these approximations
superimposed. White curves are for the assumption that $\gamma\propto\kappa^{1/2}$,
dark grey curves for $\gamma=0$, and light grey ones for $\gamma=\kappa$.}

\table{1}{S}{\bf Table 1. \rm Properties of the four cluster models used
for some of the Figures. The richness measure $C_{\rm ACO}$ is defined
by Mazure et al.\ (1995), the parameter $c$ is part of the heuristic
estimator described in Section 3.3.2, $z_{\rm d}$ and $z_{\rm s}$ are the
redshifts of the lens and the background galaxies respectively, whereas the
cosmological parameter $\sigma_8$ determines how evolved the clusters are.}
{\halign{%
\hfil#\hfil\hskip6pt&\hfil#\hfil\hskip6pt&\hfil#\hskip6pt&\hfil#\hskip6pt&\hfil#\hskip6pt&\hfil#\hskip6pt&\hfil#\hskip6pt&\hfil#\hskip6pt&\hfil# \cr
\multispan{9}\hrulefill\cr
\noalign{\vfilneg\vskip -0.4cm}
\multispan{9}\hrulefill\cr
\noalign{\smallskip}
no&$\sigma_8$&Mass&$\sigma_{\rm v,l.o.s.}$\hfil &
 C$_{\rm ACO}$ \hfil& $z_{\rm d}$\hfil& $z_{\rm s}$\hfil& $c$\hfil& \cr
  & & $[10^{15}{\rm M}_\odot]$ & [km s$^{-1}$] \hfil & & & & & \cr
\noalign{\smallskip}
\multispan{9}\hrulefill\cr
\noalign{\smallskip}
  9 & 0.54 & 0.89 & 1061 & 109 & 0.2 & 0.8 & 0.85 &   \cr
 13 & 0.63 & 1.58 & 1140 & 153 & 0.4 & 2.0 & 0.80 &   \cr
 41 & 0.79 & 2.67 & 1591 & 174 & 0.2 & 0.8 & 0.76 &   \cr
 41 & 0.92 & 2.91 & 1563 & 169 & 0.4 & 2.0 & 0.70 &   \cr
\noalign{\smallskip}
\multispan{9}\hrulefill\cr
\noalign{\vfilneg\vskip -0.4cm}
\multispan{9}\hrulefill\cr
}}

\subsection{Heuristic estimators}

\tx A typical cluster will be aspherical and
clumpy, which means that the simple approximations described above
will not hold. In fact, an exact local relation between the lens
convergence and shear is not expected to exist when the cluster is
not spherical, as the convergence will generally have a strong
dipole moment, while the shear is dominated by a quadrupole component. 
So we need to find a simple function which relates $\kappa$ and $\gamma$
{\it on average}.

\subsubsection{Numerical cluster models}

\tx A fruitful approach is to look at numerical cluster models, and find
a local relation between $\kappa$ and $\gamma$ by looking at
scatter-plots of these two quantities from the pixels of the convergence
and shear maps of model clusters.
This is useful only when we look at a fair sample of clusters models,
which is representative for the variaty of clusters found on the sky.
For this purpose we use the catalogue of high-resolution cluster models
of van Kampen \& Katgert (1997), which was constructed to mimic an observed
sample (Mazure et al.\ 1995; Katgert et al.\ 1995). 

The individual cluster models were built using a dissipationless N-body code
which was supplemented with a recipe for galaxy formation and merging
(van Kampen 1994, 1997), which makes it possible to get Abell cluster
properties like richness. A groups of particles that collapses into a
virialized group with a mass corresponding to that of a galaxy halo is,
during the simulation run, replaced by a single, massive `galaxy particle'.
However, brightest cluster galaxies like cD's and gE's are {\it not}\
replaced by single particles. For the lensing properties of their parent
cluster this is important, as both the core and the substructure of
the cluster should be modelled with sufficient resolution (Bartelmann \&
Weiss 1994; Bartelmann, Steinmetz \& Weiss 1995). We adopted a Plummer
softening parameter of $40 h^{-1}$ kpc (comoving), which is adequate for
our purposes; see van Kampen (1994) for a more comprehensive discussion
on resolution issues connected to the numerical simulation technique.
Note that the resolution of the {\it projected}\ density distribution will
automatically be higher. Therefore, more important is the use of adaptive
window smoothing (see below), which retains that resolution as
much as possible during the smoothing which is necessary for the
calculation of the lens properties, since we use the thin lens approximation.

We use 29 cluster models that constitute a complete sample
for richness $C_{\rm ACO}>75$ (the entries in boldface in Table 1
of van Kampen \& Katgert 1997). Please refer to Mazure et al.\ (1995) for
the definition of the richness parameter $C_{\rm ACO}$.
The 29 clusters were simulated for the standard
Cold Dark Matter scenario, and have $\sigma_8=0.63$ when put at a redshift
$z_{\rm D}=0.4$. However, we have also selected four specific
cluster models, with a range in mass, $\sigma_8$, and other properties
relevant for lensing, for demonstrating the various methods and tests.
The least massive of these models we may consider to be a `weak' lens,
while the most massive one is a `strong' lens with both caustic lines
present for most source redshifts.
Some of the properties of these four models, and the redshift they are
put at, are listed in Table 1. The cluster number relates to the
entry in the catalogue of cluster models of van Kampen \& Katgert (1997),
where more properties of these models can be found.

\subsubsection{Obtaining the lens properties of simulated galaxy clusters}

\tx We obtain the surface mass density (and thus the convergence) from
the numerical models using adaptive window smoothing (Silverman 1986), with
an initial (Gaussian) smoothing length of $0.25h^{-1}$Mpc. This results in
having a smoothing length of $0.05h^{-1}$Mpc in the centre of the cluster
models (which is identical to that used by Bartelmann \& Weiss (1994) for
their cluster models), and $1.0h^{-1}$Mpc in the outskirts. This
provides sufficient resolution for the lens mapping. For example, giant
arcs are formed as expected (van Kampen 1996).
The lens shear and magnification are calculated using the thin lens approximation,
as outlined in Section 2.1, with the convolution in eq.\ (\the\alphakappa)
done by Fast Fourier Transforms. All lens properties
are calculated on a 1024$\times$1024 grid which measures
4$h^{-1}$ Mpc on a side. As an example we show these maps for the fourth
cluster of Table 1 in Fig.\ \the\Flensex, along with its caustics.

In Fig.\ \the\Fkapgamex\ we plot, for the cluster models listed
in Table 1, the absolute magnification $|\mu|$ versus the
convergence $\kappa$, and the lens shear $\gamma$ versus $\kappa$,
as scatter plots. For clarity we plot just one out of every 250
pixels for each calculated map. We use these scatter plots to look
for approximate relations between the lens properties.

\subsubsection{$\gamma\propto\kappa^{1/2}$}

\tx The heuristic approach should preferrably lead to simple
approximations which can be applied to observed data in an unambiguous way.
There will be two caustic lines, which means that two parities have to
be set, so the approximation should preferrably have only a small range
of $\mu$ for which parities need to be set.

Studying Figs.\ \the\Fkapgamex, one gets the impression that, on average,
$\gamma$ is proportional to $\kappa^{1/2}$.
This assumption leads to a well-behaved relation between $|\mu|$ and
$\kappa$, which is also invertible. In general, if we assume
\newcount\gamkap \gamkap=\eqnumber
$$\gamma=[(c+c^{-1}-2)\kappa]^{1/2}\ , \eqno(\neweq)$$
with $0<c<1$ (by convention), then 
\newcount\muinv \muinv=\eqnumber
$$\mu^{-1}=(\kappa-c)(\kappa-c^{-1})\ . \eqno(\neweq)$$
This implies that there are two $\kappa$'s for each $\mu$,
but as we measure $N/N_0$, we can only obtain $|\mu|$, which can
correspond to four different values of $\kappa$.

The estimator for $\kappa$, given $|\mu|$, is then
$$\kappa_c = {c+c^{-1}\over2} - {\cal S}
 \Bigl[\Bigl({c+c^{-1}\over2}\Bigr)^2 - {\cal P}|\mu|^{-1}-1\Bigr]^{1/2}\
	, \eqno(\neweq)$$
where ${\cal P}$ is the lens parity, i.e.\ the sign of $\mu$, while
the new parity ${\cal S}$ indicates which side of the minimum we are:
it is the sign of $\kappa_{\rm min}-\kappa$, and switches sign around
$\mu=\mu_{\rm min}$. These two minima are
$$\kappa_{\rm c,min}=(c+c^{-1})/2\ ,\ 
  \mu_{\rm min}=\Bigl({c+c^{-1}\over 2}\Bigr)^2 -1 \ . \eqno(\neweq)$$
Note that $\mu_{\rm min}$ is a local {\it maximum}\ for $|\mu_{\rm min}|$.
Also, one recovers the $\gamma=0$ approximation when $c=1$.

The two critical lines are at $\kappa=c$ and $\kappa=c^{-1}$, as is
obvious from eq.\ (\preveq{3}). We compare these critical lines to
the true critical lines in Fig.\ (\the\Fapprox), for a fairly
massive cluster with central $\kappa$ larger than one.
For comparison, the critical lines corresponding to the $\gamma=0$ and
$\gamma=\kappa$ estimators are shown as well, in dark grey and light
grey colours respectively.

Combining Eqs.\ (\the\defgamma), (\the\defkappa) and (\the\gamkap), we
see that this approximation corresponds to assuming 
$$ A_{12} A_{21} = {(c-1)^2\over 2c}
   \Bigl(2-A_{11}-A_{22}\Bigr) 
   -{1\over 4}(A_{22}-A_{11})^2\ . \eqno(\neweq)$$
In order to solve this equation, we need to make a further assumption
for $A_{11}$, $A_{22}$, and $A_{12}$ (which is equal to $A_{21}$).

We can assume spherical symmetry, i.e.\ $\kappa=\kappa(x)$, but
unfortunately, as shown in Appendix A, there exists no spherical
solution for which $\gamma=[(c+c^{-1}-2)\kappa]^{1/2}$.
However, the Plummer potential, which can be written as
$$\phi(\theta)=\phi_0 \ln(\theta_{\rm c}^2+\theta^2)\ , \eqno(\neweq)$$
where $\phi_0=M\theta_{\rm c}^2/2\pi R_{\rm c}^2 \Sigma_{\rm cr}$, 
does show this behaviour for $\theta>\theta_{\rm c}$, where $\theta$ is
the angular distance from the centre of the cluster, $\theta_{\rm c}$ the
angular core radius, $\Sigma_{\rm cr}$ the critical surface mass density as
defined in eq.\ (\the\defcritden), and $R_{\rm c}=\theta_{\rm c}D_{\rm L}$
its corresponding physical core radius, where $D_{\rm L}$ is the angular
diameter distance from the lens to the observer. For this potential
$$\kappa(\theta)=2\phi_0
	{\theta_{\rm c}^2\over(\theta^2+\theta_{\rm c}^2)^2}\ \ ,\ {\rm and}\ 
  \ \gamma(\theta)=2\phi_0
	{\theta^2\over(\theta^2+\theta_{\rm c}^2)^2}\ \eqno(\neweq)$$
(Kochanek and Blandford 1991), which gives the following relation between
$\kappa$ and $\gamma$:
$$\gamma = {\kappa_0\kappa}^{1/2}-\kappa
	 = {\kappa}^{1/2}({\kappa_0}^{1/2}-{\kappa}^{1/2})\ , \eqno(\neweq)$$
where $\kappa_0=\kappa(0)=2\phi_0/\theta_{\rm c}^2$.
For small $\kappa<\kappa_0$ we then have
$\gamma\approx(\kappa_0\kappa)^{1/2}$, so we can identify
$\kappa_0=(c+c^{-1}-2)^{1/2}$. Clearly, we need to supplement this
potential with extra depth in the core region, as $\kappa_0$ for the
Plummer model will not be very large for typical values of $c$.

One might be able to get $\gamma\propto\kappa^{1/2}$ by constructing
more complicated potentials, involving an elliptical component with
ellipticity growing as a function of radius (constant ellipticity
does not work), or a quadrupole component of some sort. However, we
can simply treat this approximation as a heuristic one, motivated
by simplicity and invertability.

In order to obtain the convergence and shear from the numerical models,
which are discrete in nature, we had to apply smoothing. Even though
the adaptive smoothing allows relatively high resolution to be retained
in the core of the cluster, one might worry that the central values of
both $\kappa$ and $\gamma$ are artificially reduced. Because
the shear is a global function of the convergence through the
convolution of eq.\ (\the\alphakappa), this would be most severe for
$\kappa$.  We therefore tried several basic smoothing lengths, from
$0.1h^{-1}$Mpc to $0.5h^{-1}$Mpc. The ones smaller than the value of
$0.25h^{-1}$Mpc that we actually use gave the same results,
i.e.\ similar scatter plots and the same value for $c$ from the fit.
The only difference is in the very centre of the cluster, where
$\kappa$ is slightly more peaked so that $\gamma$ has somewhat larger
maximum values.
However, the discreteness of the numerical simulation does becomes
quite visual for these small values of the basic smoothing radius.
Oversmoothing affects the results more severely, pushing $\gamma$ up in
the outskirts and down in the centre. So, provided that the basic
smoothing length is chosen sensibly, it seems that the use of adaptive
window filtering results in reliable convergence, shear and
magnification maps.

\subsubsection{$\gamma\propto\kappa$}

\tx Another assumption which leads to a simple, invertible relation
which has $\mu(0)=1$, is $\gamma=a\kappa$. This leads to
$$\mu^{-1} = (\kappa-1)^2 - (a\kappa)^2\ ,  \eqno(\neweq)$$
and inverts as
$$\kappa_{\rm a}=1-{\cal T}\Bigl[(1-a^2){\cal P}|\mu|+a^2\Bigr]^{1/2}\ , 
  \eqno(\neweq)$$
where ${\cal T}$ is a parity similar to the parity ${\cal S}$ of the
$\gamma\propto\kappa^{1/2}$ approximation.
This approximation corresponds to pivoting the $\gamma=0$ approximation
around $(0,0,1)$ in Fig.\ (\the\mukapgam). It can serve as an
approximation intermediate to the $\gamma=0$ and $\gamma=\kappa$
approximations, but it has the disadvantage of the extra parity.
More importantly, this approximation is a bad fit to the
numerical simulations, so we will refrain from using it.

\newfig \newcount\Fshear \Fshear=\fignumber
\figps{\fignumber}{D}{
\psfig{file=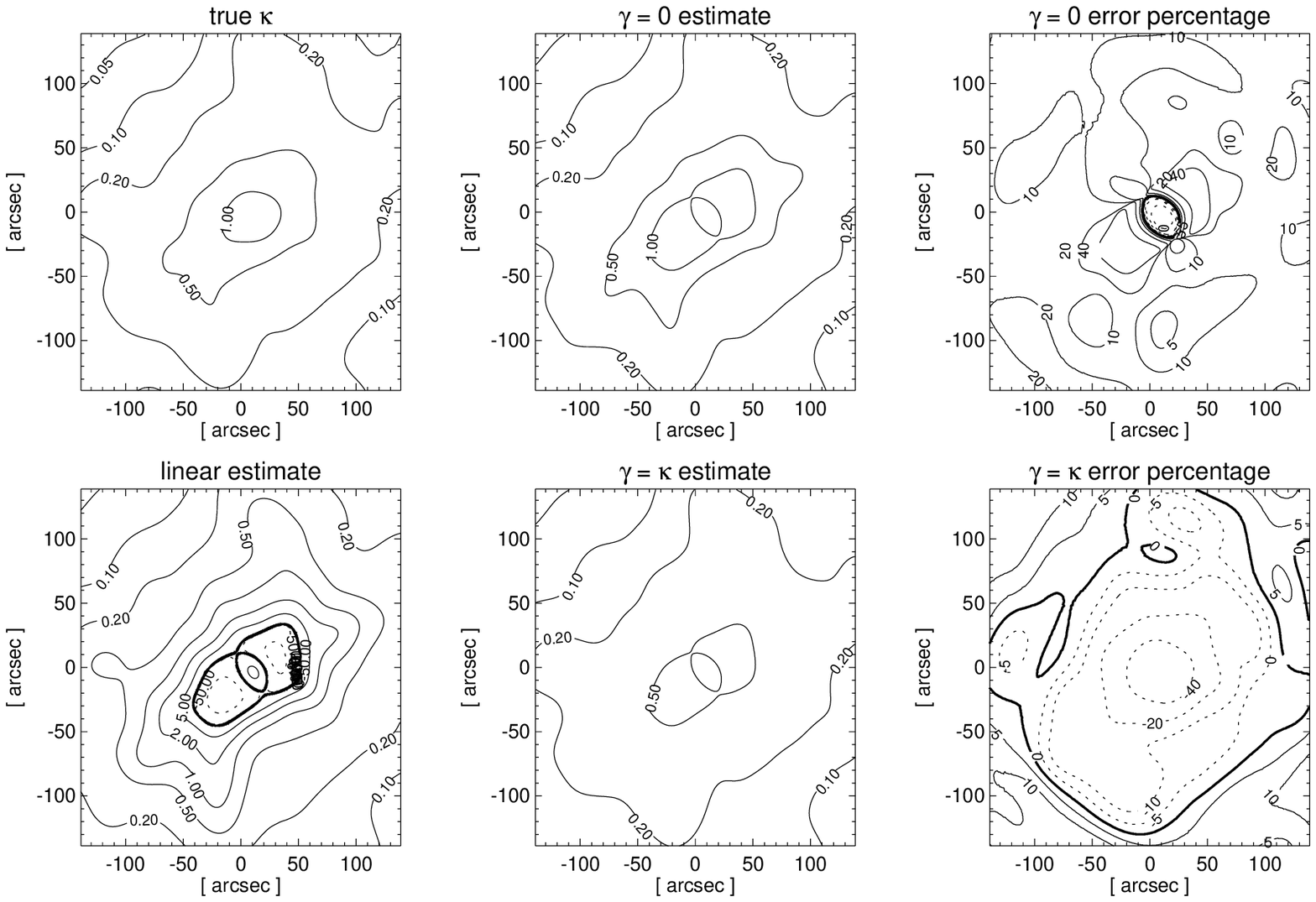,width=16.0cm,silent=}
\psfig{file=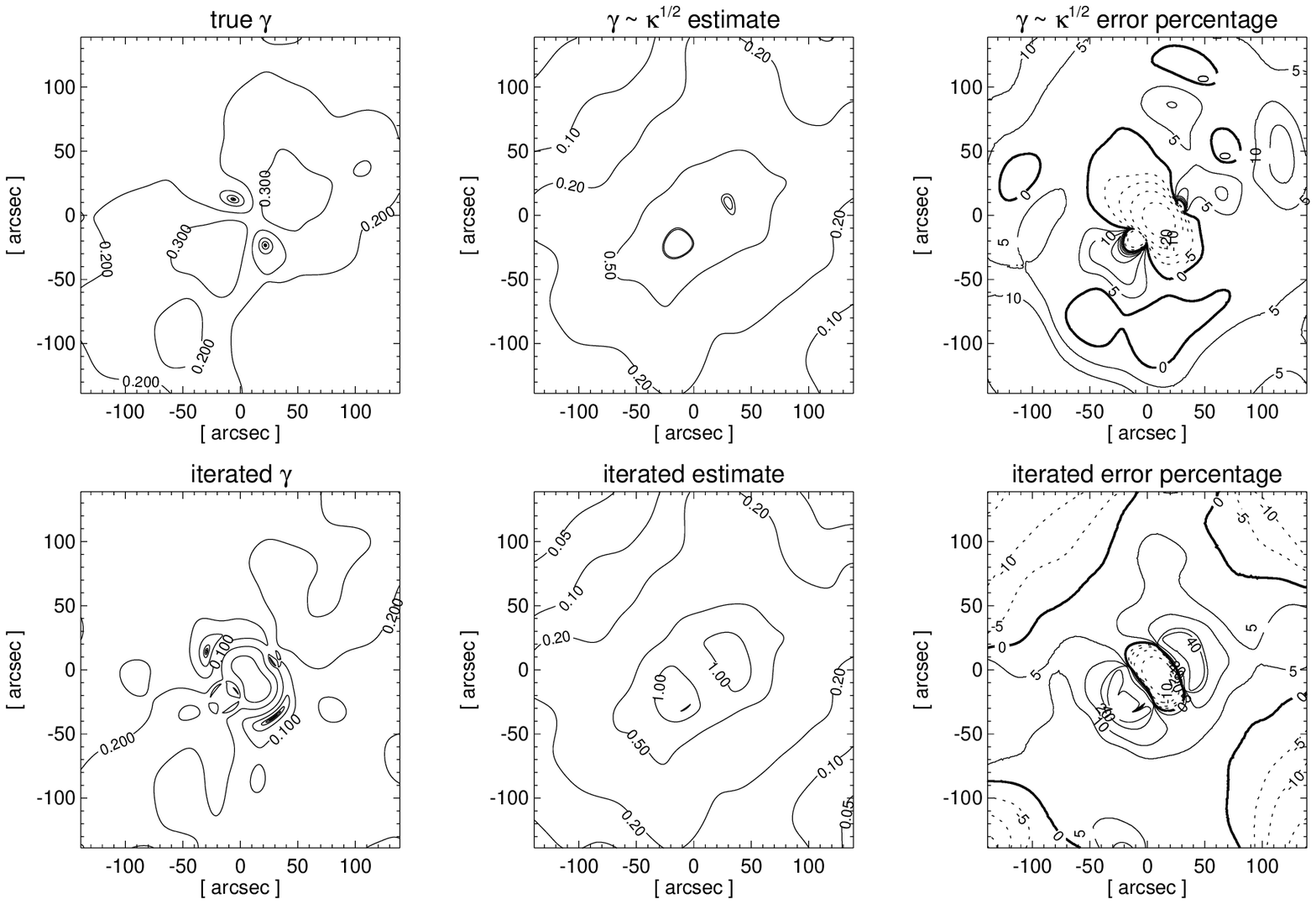,width=16.0cm,silent=} }
{{\bf Figure \the\fignumber.} Comparison of estimated $\kappa$ versus
true $\kappa$. The middle column shows estimated $\kappa$ maps
for the $\gamma=0$, $\gamma=\kappa$, $\gamma\propto\kappa^{1/2}$,
and iterative (1 step) estimators. The right column shows the fractional
difference (as percentages) of true versus estimated $\kappa$ maps.
The top left panel shows the true $\kappa$, the panel below that the
linear estimate for $\kappa$, i.e.\ ($\mu$-1)/2.
The iteratively estimated shear is shown in the bottom left corner,
below the true shear.
The model cluster used is the fourth entry of Table 1.}

%

\subsection{Iterative estimate}

\tx For each of the convergence estimators discussed above, we can find an
iterative solution by calculating the shear corresponding to the estimate
for $\kappa$ using the thin lens approximation described in Section 2.1.
In other words, we use the estimate for $\kappa$ to calculate the
lens deflection ${\bf\alpha}$ using eq.\ (\the\alphakappa),
and then get $\gamma$ from ${\bf\alpha}$.
This shear is then used to find a new estimate for $\kappa$ by simply
applying eq.\ (\the\mukapgam). 
Because of the convolution, this estimate is non-local.

A problem with this estimate is that the magnification observed is usually
smoothed on a scale larger than the smallest significant structures in the lens.
The shear field has a strong quadrupole component, with small scale structure
which is not present in the measured smoothed magnification, although it
is there in the unsmoothed magnification.
This means that the new estimate for $\kappa$ derived from eq.\
(\the\mukapgam) will have this quadrupole structure imprinted by the shear
estimate. In the next step the second estimate for the shear will have even
more structure on small scales, generated by the small scale structure which
is present in the second estimate for $\kappa$. So convergence is not
achieved.


\subsection{Comparison}

\tx In principle, the best strategy, if data quality allows it, is to use
the iterative method, starting from one of the estimates discussed above.
If parities can be assigned reliably, $\kappa_{\rm c}$ seems a good choice,
otherwise one should start from $\kappa_1$, when one has only approximate
knowledge about one critical line.
However, because of the problems associated with the iterative estimate
due to the likely smoothness of the observed lens magnification, in practice
the $\kappa_{\rm c}$ estimator is to be preferred.

We show the caustic lines corresponding to the one-caustic
and $\kappa_{\rm c}$ estimators for two of the lens models in
Fig.\ (\the\Fapprox), along with the real caustic lines for those
models. Clearly the $\kappa_{\rm c}$ estimators has its caustic lines
closest to the real ones. The one-caustic estimators are mostly useful
for reference, as possible limits, of for observational data of poor
quality. A comparison of caustics is not enough to compare the usefulness
or goodness of the various approximations. The $\kappa_{\rm c}$ and
$\kappa_1$ estimators, for example, cross at $\kappa=c+c^{-1}-2$, so a
total mass estimation where the total mass is calculated beyond this
crossover point might be similar for both approximations.
We therefore test all estimators in the next Section, both on perfect
magnification data as well as on mock galaxy counts.

\section{Testing the convergence estimators on cluster models}

\tx We test the convergence estimators on a sample of numerical cluster
models in order to examine how well each of them performs in different
circumstances. The models allow us to test how well one can
reconstruct the convergence from the magnification alone, by simply
comparing the result of applying the estimator to the true convergence.
With the addition of simple background galaxy distributions we
also produce mock lensed galaxy distributions which we can 'observe',
thus providing a direct test of the magnification method. Such tests
should reveal intrinsic uncertainties and systematic offsets when this
method is applied to observational data.

We perform two types of tests on two types of data. We test on
clean simulated magnification maps with full knowledge of the 
image parity, and for a single source redshift, and we test on
mock lensed galaxy distributions, performing the whole route from
galaxy counts to mass estimation.

\subsection{Estimated versus true convergence for known magnification}

\tx At this point we forget about all possible observational problems.
In other words, we assume that we have a perfect magnification map of the
lensing cluster (including the parity), and investigate how well the
approximations allow us to reconstruct the convergence from just this
magnification map. This should show us the best possible result each
approximation can provide us, and reveals systematic effects associated
with the estimators alone (as compared to other possible, mainly
observational, sources of error).

In order to quantify these statements, we look at statistics which are
based on a pixel-to-pixel comparison of estimated convergence maps to
true convergence maps for a sample of rich cluster models. We are
interested in the average performance for a typical rich cluster. For this
purpose we selected 29 rich cluster models (see Section 3.3) from the
catalogue of van Kampen \& Katgert (1997). We then calculated the average
of various statistics applied to each of these models. The results are
listed in Table 2, and are described below.

For each model we calculated the
minimum and maximum difference between the estimated and the true pixels/bins,
denoted in Table 2 by `min' and `max' respectively, and the average over
all pixels of this difference, denoted by `mean'. These statistics describe
local and mean deviations from the true convergence. Two more statistics
describe the scatter in the residual of true minus estimated convergence:
`abs' is just the average absolute deviation, while `rms' is the standard
{\it r.m.s.}\ deviation. These are good measures for the accuracy of the
estimators. All statistics were calculated for absolute differences in
$\kappa$, and for fractional differences, expressed in percentages.

An additional estimator has been added to Table 2, labeled `mean', which
just comprises taking the mean of the $\gamma=0$ and $\gamma=\kappa$
estimators. This estimator has no physical basis, and even produces
imaginary shear for some values of $\kappa$, but might be useful in
practice as the $\gamma=0$ estimator tends to overestimate and the
$\gamma=\kappa$ tends to underestimate.

\subsubsection{Maps}

\tx Using the thin lens approximation, as summarized in Section 2.1,
we produce maps of the lens convergence, shear and magnification for
the selected four cluster models. We then use the magnification only
(with full knowledge of its parity, though), to reconstruct convergence
maps using the various assumptions about the shear, and compare these
to the true convergence. This is shown in Fig.\ \the\Fshear\ for the
most massive model cluster, as this one has the largest range of possible
$\kappa$ values and therefore provides the most stringent tests for the
estimators. Note that this clusters has a value of $\sigma_8$ which is
far larger than allowed for the standard CDM scenario (e.g.\ van Kampen
\& Katgert 1997 and references theirin).

In Fig.\ \the\Fshear\ we see that linear estimator performs very poorly,
as it just follows the magnification, including the caustics. It is
only doing well for small $\kappa$, as expected. The $\gamma=0$
assumption produces an overestimate for the convergence for all regions
of the cluster. The $\kappa=\gamma$ estimator underestimates the mass
in the central regions of the cluster, and (slightly) overestimates
for $\kappa<0.2$. The $\gamma\propto\kappa^{1/2}$ estimator clearly
performs best, even better than the iterative estimator. 

As the cluster shown in Fig.\ \the\Fshear\ is fairly extreme, we should
consider the performance statistics listed in Table 2 that were obtained
for the rich cluster sample, which contains less evolved clusters with
smaller overall convergences. Much of what was seen in Fig.\ \the\Fshear\
is now expressed quantitatively. The linear estimator diverges, while the
$\gamma=0$ and $\gamma=\kappa$ estimates generally over- and underestimate,
respectively. Note that the fractional and absolute statistics weight pixels
differently, the former giving more weight to the more noisy pixels outside
of the cluster, and the latter to the central regions. This results in a
mean overestimate in the fractional statistics of both the $\gamma=0$ and
$\gamma=\kappa$ estimators.
The $\gamma\propto\kappa^{1/2}$ and iterative estimators perform best,
although the improvement over the other estimators is certainly not
dramatic.

\subsubsection{Radial profiles}

\tx Besides testing the estimators on the full 2D map of the lens
magnification, which is hard to obtain in practice, we test them
on azimuthally averaged `magnification profiles' as well. These have
already been obtained observationally (Fort et al.\ 1997; Taylor
et al.\ 1998), and are therefore of particular interest.

Again we first look at radial binning of the magnification map
directly obtained from a numerical simulation, i.e.\ we only
investigate the effect of the annular binning procedure.
We performed the same statistics on the same model cluster sample as
above, but now on the radially binned data. One expects that the
binning reduces the scatter in the local relations, but at the same
time binning means loss of information, especially for clumpy
aspherical clusters. 

From the performance statistics listed in the lower half of Table 2
we can see that the perfomance for profiles is worse than for the maps.
This stems from the fact that a lot more weight is put on the central
region of the clusters due to the annular binning, where the uncertainties
are largest. The $\gamma=0$ and $\gamma\propto\kappa^{1/2}$ approximations
clearly perform best in terms of accuracy, as expressed by the {\it r.m.s.}\
statistic, whereas the mean of the $\gamma=0$ and $\gamma=\kappa$ estimator
has the smallest systematic error. 

\table{2}{S}{\bf Table 2. \rm Performance statistics of the convergence
estimators, based on a pixel-by-pixel comparison of the estimated versus
true convergence maps (top half), and a bin-to-bin comparison for the
estimated versus true profiles. A positive sign corresponds to overestimating
the convergence. Please refer to the main text for a description of
the statistics.}
{\halign{%
#\hfil\hskip6pt&\hfil#\hskip6pt&\hfil#\hskip6pt&\hfil#\hskip6pt&\hfil#\hskip6pt&\hfil#\hskip6pt&\hfil#\hskip6pt&\hfil#\hskip6pt&\hfil# \cr
\multispan{7}\hrulefill\cr
\noalign{\vfilneg\vskip -0.4cm}
\multispan{7}\hrulefill\cr
\noalign{\smallskip}
Statistic & linear & $\gamma=0$ & $\gamma=\kappa$ & mean & 
  $\gamma\propto\kappa^{\scriptstyle{1\over2}}$ & Iterat. \cr
\noalign{\smallskip}
\multispan{7}\hrulefill\cr
\noalign{\smallskip}
\multispan{7}\centerline{Absolute statistics for 2D maps}\cr
\multispan{7}\cr
     min &  -0.01 &  -0.09 &  -1.19 &  -0.57 &  -0.43 &  -0.12 \cr
     max &   7.27 &   0.23 &   0.06 &   0.10 &   0.06 &   0.11 \cr
    mean &   0.11 &   0.01 &  -0.02 &  -0.01 &  -0.02 &  -0.01 \cr
     abs &   0.11 &   0.01 &   0.03 &   0.02 &   0.03 &   0.01 \cr
     rms &   0.46 &   0.03 &   0.09 &   0.04 &   0.04 &   0.01 \cr
\multispan{7}\cr
\multispan{7}\centerline{Fractional statistics for 2D maps}\cr
\multispan{7}\cr
     min &  -6.04 &  -6.09 &  -6.50 &  -6.13 &  -6.93 &  -8.32 \cr
     max &  21.53 &   1.97 &   1.37 &   1.64 &   0.75 &   0.70 \cr
    mean &   0.68 &  -0.49 &  -0.83 &  -0.66 &  -1.58 &  -1.44 \cr
     abs &   1.87 &   0.97 &   1.04 &   0.96 &   1.61 &   1.47 \cr
     rms &   2.72 &   1.24 &   1.17 &   1.18 &   1.18 &   1.52 \cr
\multispan{7}\cr
\multispan{7}\centerline{Absolute statistics for radial profiles}\cr
\multispan{7}\cr
     min &   0.05 &  -0.20 &  -6.44 &  -2.74 &  -1.78 &   --   \cr
     max & 864.70 &   1.78 &   0.11 &   0.26 &   0.12 &   --   \cr
    mean &  20.02 &   0.33 &  -0.72 &  -0.19 &  -0.32 &   --   \cr
     abs &  20.02 &   0.35 &   0.78 &   0.34 &   0.34 &   --   \cr
     rms &  86.38 &   0.44 &   1.65 &   0.69 &   0.44 &   --   \cr
\multispan{7}\cr
\multispan{7}\centerline{Fractional statistics for radial profiles}\cr
\multispan{7}\cr
     min &   2.40 &   0.06 & -11.35 &  -4.79 &  -4.97 &   --   \cr
     max &1121.60 &   4.60 &   1.80 &   2.74 &  -0.92 &   --   \cr
    mean &  37.61 &   2.26 &  -1.39 &   0.44 &  -2.90 &   --   \cr
     abs &  37.61 &   2.35 &   2.68 &   1.79 &   3.01 &   --   \cr
     rms & 116.44 &   1.23 &   3.73 &   2.02 &   1.11 &   --   \cr

\noalign{\smallskip}
\multispan{7}\hrulefill\cr
\noalign{\vfilneg\vskip -0.4cm}
\multispan{7}\hrulefill\cr
}}

\newfig \newcount\Flenscount \Flenscount=\fignumber
\figps{\fignumber}{S}{
\psfig{file=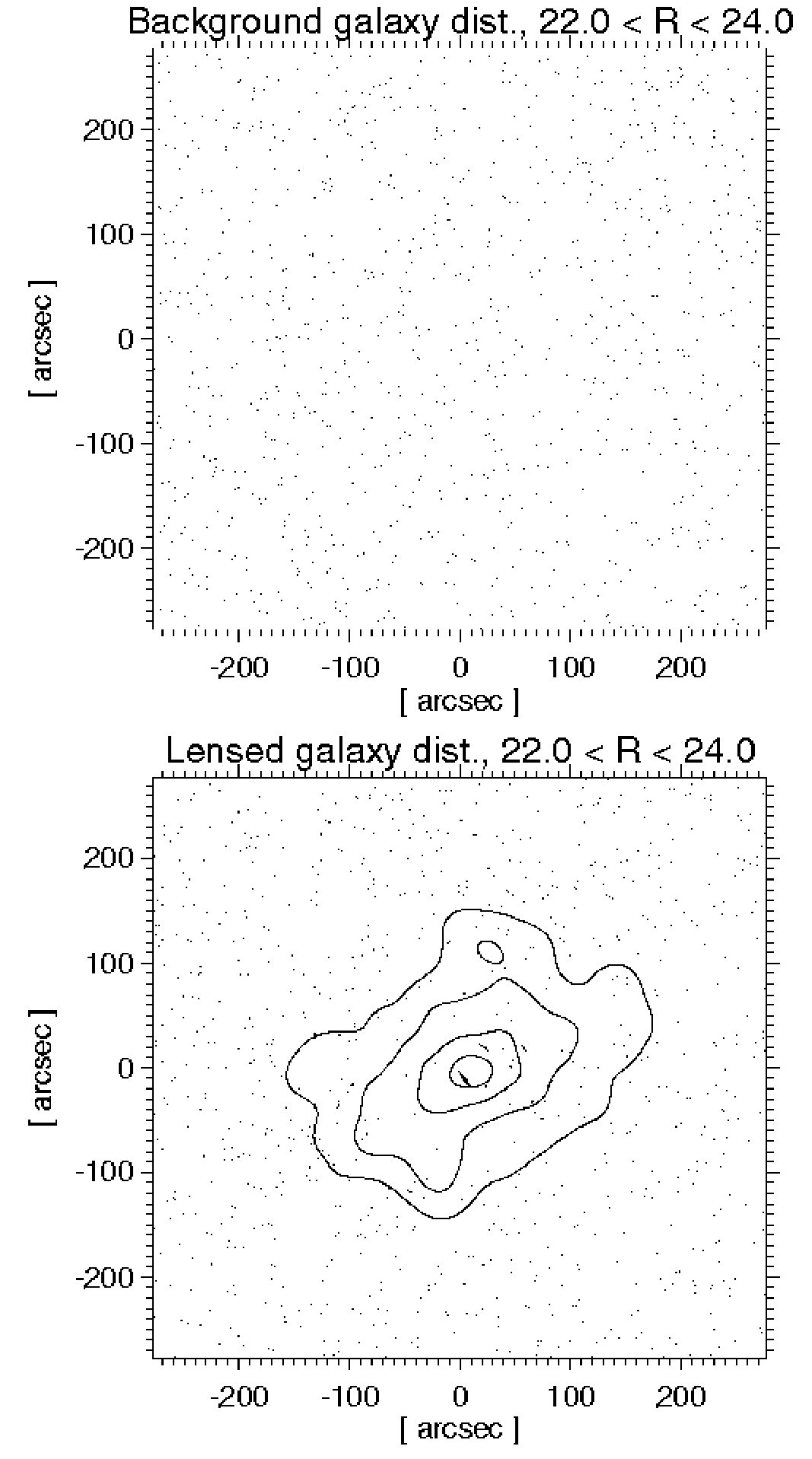,width=8.5cm,silent=} }
{{\bf Figure \the\fignumber.} Example of a number count simulation.
The top panel shows a generated background
galaxy distribution using the simple model of BTP. The bottom panel
shows the same distribution lensed, with the true convergence contours
superimposed (for $z_{\rm s}=0.8$).
Galaxies are generated up to $R=25$, but only plotted in the range $22<R<24$.
This allows for a factor of 2.51 magnification, which is sufficient in
most cases.}

\newfig \newcount\Ffieldcountex \Ffieldcountex=\fignumber
\figps{\fignumber}{S}{
\psfig{file=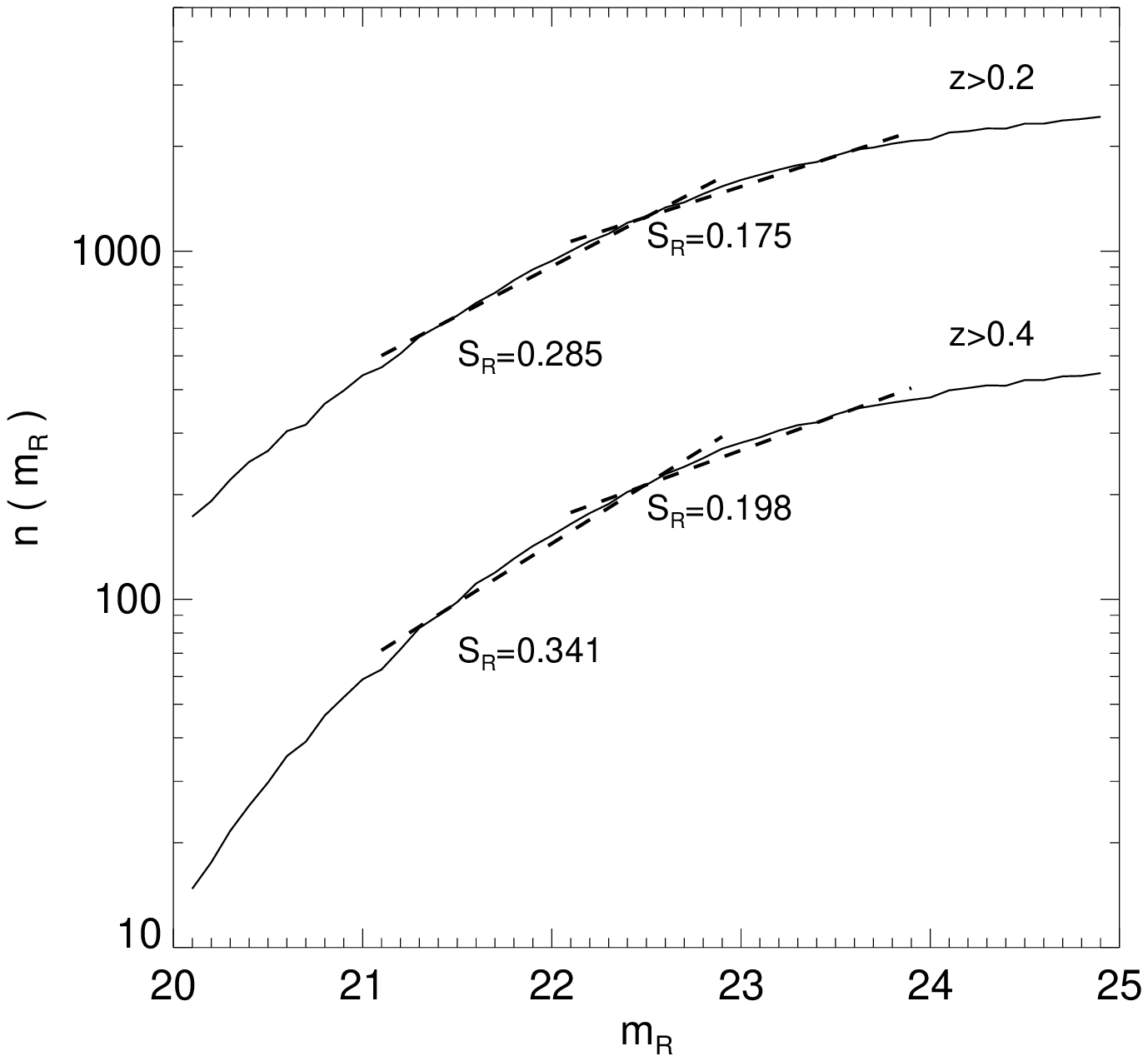,width=9.0cm,silent=} }
{{\bf Figure \the\fignumber.} Cumulative luminosity function for all 32
backgrounds stacked together, with fits for the average slope
over a range in $R$. The top curve is for galaxies beyond $z=0.4$,
the bottom one for $z>0.2$. The normalisation is arbitrary, and chosen
for clarity.}

\newfig \newcount\Fcountmapex \Fcountmapex=\fignumber
\figps{\fignumber}{D}{
\psfig{file=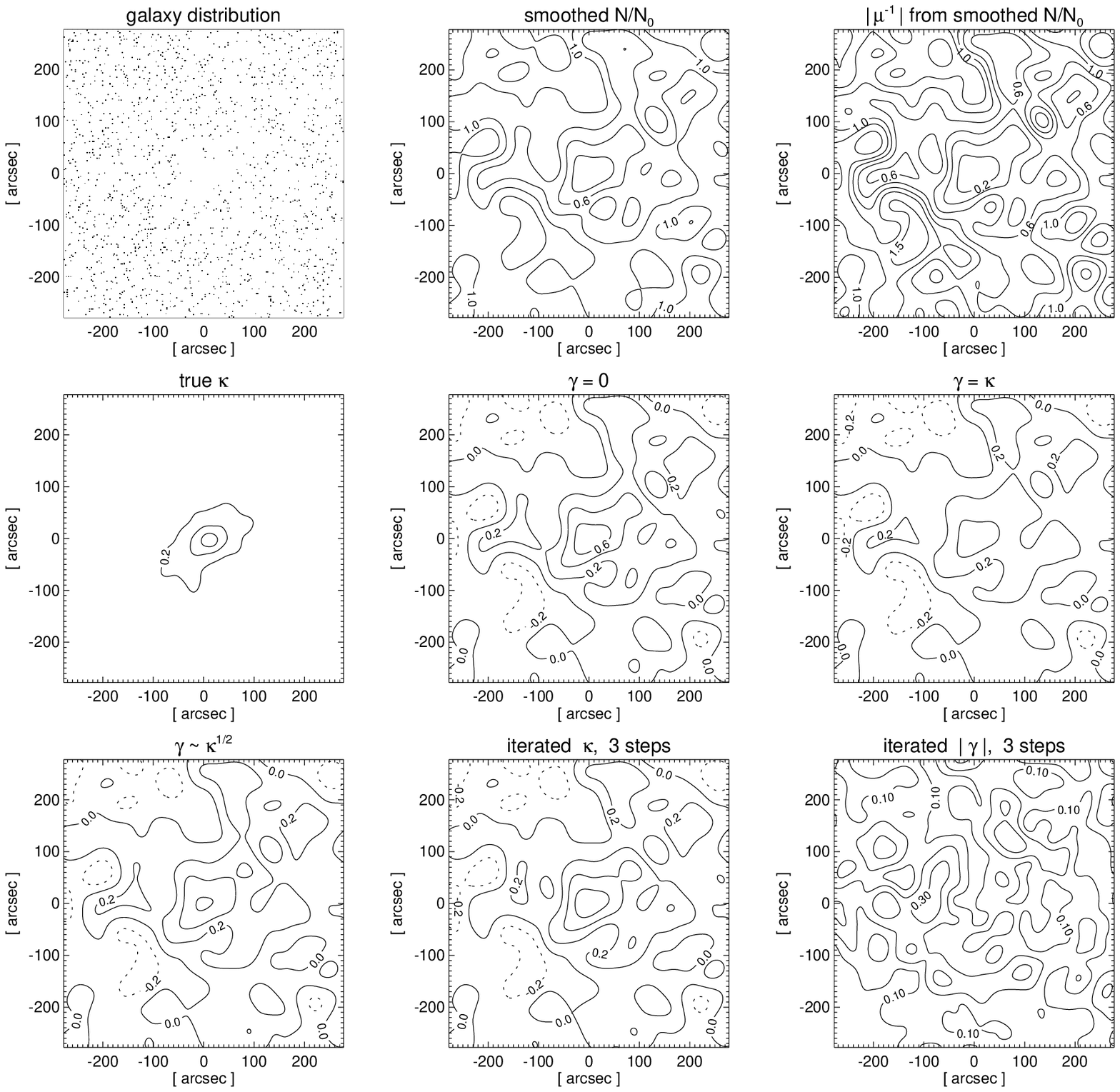,width=17.5cm,silent=} }
{{\bf Figure \the\fignumber.} Illustration of the complete route from
number counts to estimation of the lens convergence, for a fairly massive
cluster (the fourth entry from Table 1). The top left shows
the number count distribution in the range $22<R<24$, which is smoothed
as described in the main text to obtain a map of number count variations,
$N/N_0$. From this map we obtain the magnification map, shown in the
top right panel. The fourth panel shows the true convergence map,
obtained directly from the numerical simulation, which we try to estimate.
The next four panel show estimates obtained from the magnification map alone,
for, respectively, the $\gamma=0$ approximation, for $\gamma=\kappa$, for
$\gamma\propto\kappa^{1/2}$, and performing three steps of the iterative
scheme. The bottom right panel shows the shear correspinding to this
iterative estimate.}

\newfig \newcount\Fcountprofex \Fcountprofex=\fignumber
\figps{\fignumber}{S}{
\psfig{file=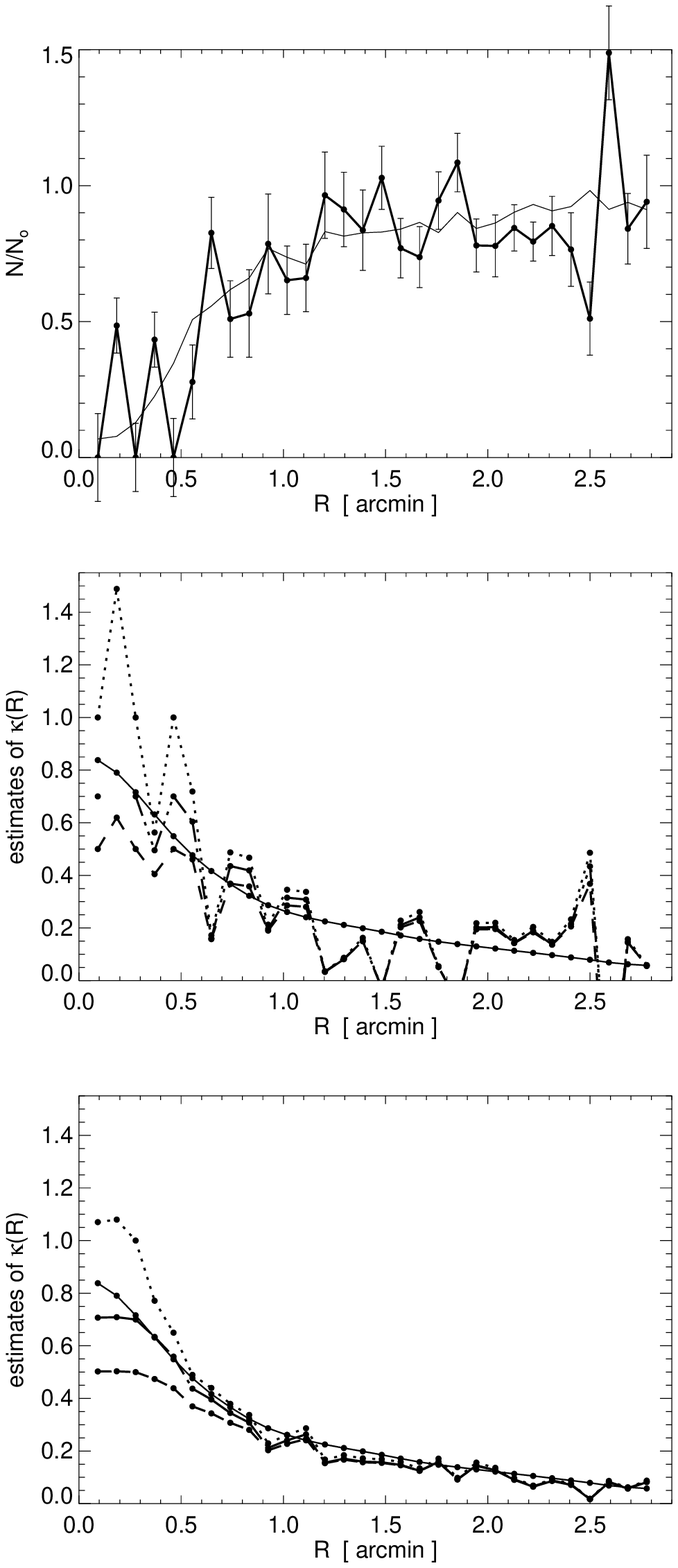,width=8.4cm,silent=} }
{{\bf Figure \the\fignumber.} Simulated number count profiles for the same
cluster used for Fig.\ \the\Fcountmapex, and also for $22<R<24$.
The top panel shows a number count
profile for a single profile (thick solid line), along with the average
of number count profiles for 32 different profiles (thin solid line).
The latter provide an estimate for the intrinsic uncertainty due to shot
noise and clustering of the background galaxies, and these are overplotted
as error bars.
The second panel shows the convergence estimates from the single background
number count profile shown in the top panel, for $\gamma=0$ (dotted line),
$\gamma=\kappa$ (dashed line), and $\gamma\propto\kappa^{1/2}$ (dot-dash line).
The true convergence profile, for $z_{\rm s}=0.8$, is plotted as a solid line.
The bottom panel shows the same, but for the average over 32 different
background populations. A intrinsic count slope of 0.198 is used.}

\newfig \newcount\Fhistmass \Fhistmass=\fignumber
\figps{\fignumber}{S}{
\psfig{file=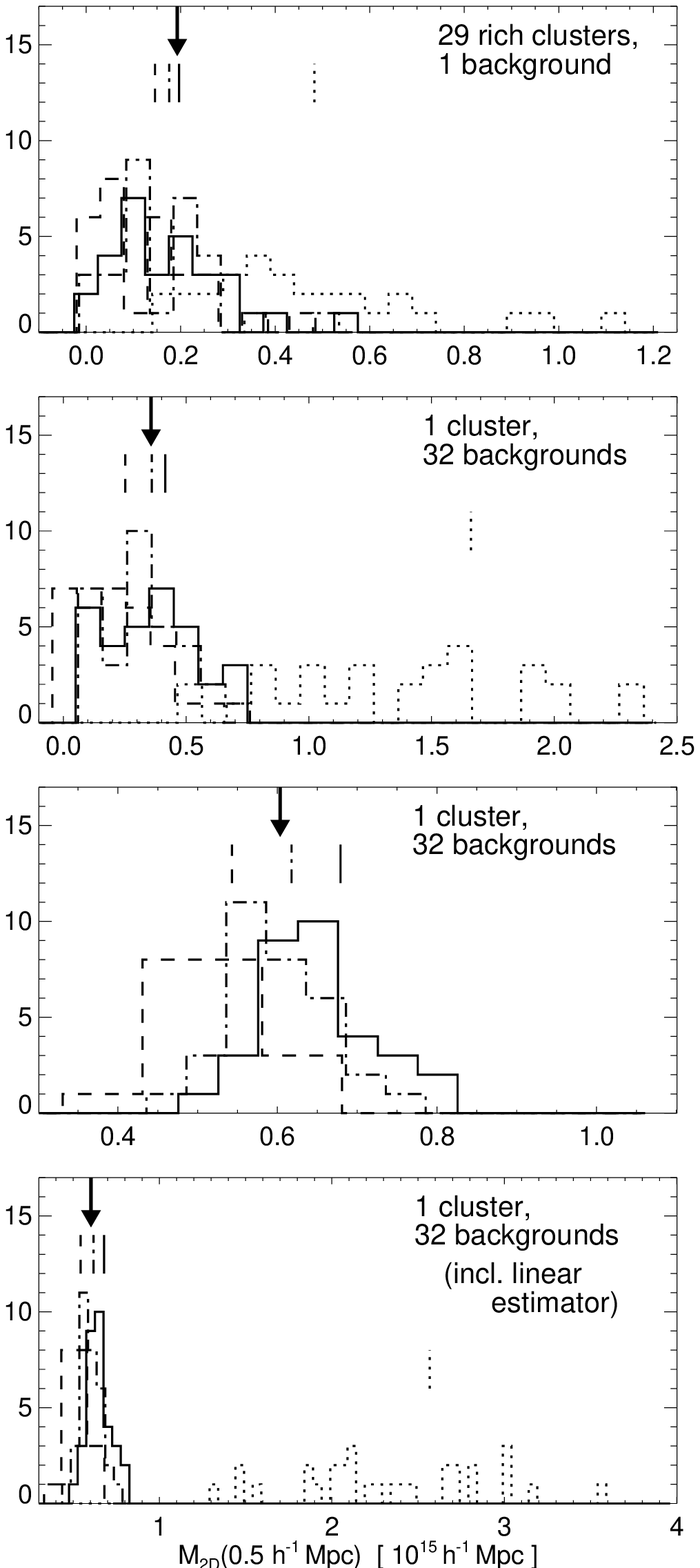,width=8.4cm,silent=} }
{{\bf Figure \the\fignumber.} Histograms of estimated total projected
cluster masses within $0.5 h^{-1}$Mpc, obtained from number count profiles.
Dotted lines deonte the linear (weak lensing) approximation, solid lines
are for $\gamma=0$, dashed for $\gamma=\kappa$, and dot-dashed for 
$\gamma\propto\kappa^{1/2}$. The top panel shows the distribution for
29 rich clusters put at $z_{\rm D}=0.4$ in front of the same background
galaxy population. The vertical lines indicate the average over the
estimated projected masses of these clusters, while the arrow show the
average over their true projected masses.
The second panel shows the same, but now for one cluster (entry three
from Table 1) in front of 32 different backgrounds. The lines and arrow
are now the estimated and true masses of that cluster. The third panel shows
the same cluster for larger $\sigma_8$, i.e.\ more massive. The fourth panel
is the same as the third, but with the linear estimator included.}

\subsection{Estimated versus true convergence for simulated observational
data}

\subsubsection{Simulated background galaxy distributions}

\tx The tests we performed so far were for the best possible data
set, a continuous magnification map or annular profile, with known
sign, and one source redshift. Obviously, this is not going to be
the case for observational data, which has many intrinsic uncertainties
as described in Section 2.3.

The main intrinsic source of error is the clustering of background
galaxies, the variation in their redshifts, and the fact that we
usually have a limited number of galaxies to count, which introduces
shot noise. In order to examine the effect of these uncertainties
on the convergence estimate, we need to construct magnification maps
derived from number counts of a clustered set of background galaxies
which are at different redshifts.

We use background galaxy distributions which were generated as in BTP,
who adopt the lognormal-Poisson model (Coles and Jones 1991) as a
simple but sufficient model for the clustering of background galaxies.
These galaxies were generated in planes of constant redshift, with
luminosities drawn from a Schechter luminosity function
$$\phi(L)=\phi_*(z) \exp(-L/L_*),\ \eqno(\neweq)$$
where $\phi_*(z)=0.02h^2(1+z)^2{\rm Mpc}^{-3}$, and $L_*$ is taken
constant. The redshift distribution for these galaxies can be well
approximated by $n(z) = 4 z_*^{-3} z^2 e^{-2 z/z_*}$, with $z_*=0.8$. 
Like BTP, we consider R band counts. In order to be
able to count lensed (i.e.\ amplified !) distributions down
to a limiting magnitude $m_{\rm lim}$, we generate background
distributions down to at least $m_{\rm lim}+1$.

The field count slopes of these distributions for two magnitude
intervals and two lens redshifts are obtained from fits to the
luminosity functions of the 32 samples generated. This is shown
in Fig.\ \the\Ffieldcountex, with the values of the slopes annotated.
These luminosity functions are good fits to the observed ones (e.g.\
Metcalfe et al.\ 1995), except possibly for the faint end, where
the slope should remain roughly constant instead of flattening down,
i.e.\ where the simple assumptions of BTP are likely to break down.
However, the observations at the faint end are still controdictionary.
Furthermore, a flattening slope actually provides a useful test for our
methods, as other bands like the U-band will show this behaviour.

With these background distributions it is straightforward to produce
mock observations. All galaxies (i.e.\ down to $R=25$) are mapped
to the image plane using the lens mapping, with the deflection angle
calculated using the thin lens approximation, as described earlier.
We then select galaxies for counting from a range in magnitude.

Before we can obtain a magnification map from these number counts,
we need to either smooth the number count distribution, or
azimuthically bin the counts to obtain count profiles. For a limited
number of galaxies this is the only feasible alternative. We
consider both alternatives, and test for the intervals $22<R<24$.
An example of the former is given in Fig.\ \the\Flenscount, which 
shows a background distribution cut to $22<R<24$ and its
corresponding lensed distribution, also cut to $22<R<24$ but obtained
using the full $R<25$ background population. Close to the caustics
the amplification will be more than one magnitude (i.e.\ $\mu>2.51$),
so we are likely to miss a few galaxies there, which will slightly
enhance the presence of the caustics in our simulations.

The mock observations do not simulate colour cutting, and problems
associated with masking the cluster galaxies. The cluster galaxies
are modelled (see van Kampen \& Katgert 1997 for details), and take
part in the lensing, but are not put in the observed image and then
masked out. These observational difficulties are hard to model,
and beyond the scope of the present paper.

\subsubsection{Setting parity and the estimator parameters}

\tx Setting the parity is the biggest problem for all estimators,
and will be the most significant uncertainty in the central regions
of the observed cluster. However, giant arcs, especially with
redshifts measured, can be used to guess were the caustics are.
In the tests we performed above the parity was determined from the
simulations, i.e.\ we actually used the sign of $\mu$ for estimating
the convergence. We now disregard any such knowledge, and set the
parity by hand.

Concerning the $\kappa\propto\gamma^{1/2}$ approximation,
we can get its parameter $c$ from observed data only if we make an
assumption for the behaviour of $\kappa$ between the critical curves.
For  number counts in radial bins we can estimate $c$ if we know both
the position of the minimum $\mu_{\rm min}$, $\theta_{\rm min}$, and
one of the critical lines (or both), $\theta_{\rm inner}$ or
$\theta_{\rm outer}$, and assume a functional form for $\kappa$ as a
function of $\theta$ within the range
$[\theta_{\rm inner},\theta_{\rm outer}]$. 
For example, the assumption that between the critical lines
$\kappa\propto \theta^{-1}$ leads to:
$$c\approx\Bigl(2{\theta_{\rm outer}\over\theta_{\rm min}}-1\Bigr)^{1\over 2}
   \approx\Bigl(2{\theta_{\rm inner}\over\theta_{\rm min}}-1\Bigr)^{-{1\over 2}}
   \approx \Bigl({\theta_{\rm outer}\over\theta_{\rm inner}}\Bigr)^{1\over 2} .
   \eqno(\neweq)$$
If all three characteristic positions are measurable,
an average of the three $c$-estimators should provide the best estimate.
We can use this estimate for $c$ to estimate $\kappa(\theta)$,
and use that to obtain a new estimate for $c$. Such an iteration should
work if we deal with the parities properly.


\subsubsection{Smoothed maps from discrete number counts}

\tx We generate mock number counts by lensing background distributions
which were produced up to the red magnitude $R=25$ as described in
Section 4.1.1. We then count galaxies in the range $22<R<24$ only.
For almost all galaxies the magnification is less than one magnitude,
so this provides a good approximation. We also count galaxies in the
same field-of-view for the unlensed distribution, in order to obtain
the field count $N_0$. The average slope $S_R$ is fitted from
a reconstructed luminosity function obtained from the same unlensed
distribution, as this depends on the redshift of the lens,
$z_{\rm D}$. For several $z_{\rm D}$'s and $R$ ranges
we show these fits for $S_R$ in Fig.\ \the\Ffieldcountex.

In the case that we have a sufficient number of background galaxies
we can try to obtain a magnification map from the number counts. We
have to deal with shot noise, and therefore obtain a smoothed number
count map from the discrete galaxy distribution. This is {\it not}\
straightforward, as we have to deal with caustics, where $\mu^{-1}=0$.
If we smooth the number counts without taking parity into account,
there will be few points on the map that even approach zero after
the smoothing operation. We therefore apply the following `trick':
we already set parity by setting $N/N_0$ negative where we believe
that $\mu^{-1}$ is negative. This ascertains that the caustics we
fix by setting the parity remain in place, and gives a much better
estimate for the magnification near the caustics as well.

The smoothing scale needed is determined by the average surface number
density of galaxies. We set it to be three times the Poissonian nearest
neighbour distance, i.e.\ $3(\pi<n>)^{-1/2} \approx 1.7 <n>^{-1/2}$.
This is to make sure the smoothing is sufficient for the core region,
which is relatively devoid of galaxies due to the effect we try to measure.

An example of estimating the convergence map from simulated number
counts is shown in Fig.\ \the\Fcountmapex. Again the fourth cluster
from Table 1 was used. As it is put at a redshift of 0.4, we used
a slope $S_R=0.197$ for obtaining the magnification map from the
number counts. We see that we can reasonably well reconstruct the 
{\it main features}\ of the convergence map from these counts, but
a lot of information is lost. The cluster is detected, most precisely
by the $\gamma\propto\kappa^{1/2}$ estimator, but the limitations of
the magnification method are obvious. Of course one can simply improve
signal-to-noise by going to fainter magnitudes, which increases the
number of galaxies and includes galaxies at higher redshift, which pushes
up the convergence.

\subsubsection{Annular binning of number counts}

\tx Usually the number of background galaxies is not sufficient to
produce 2D magnification maps, and one is restricted to counts in
annular bins. We again only select galaxies with $R$ magnitudes in
the range $22<R<24$, and count them in bins around the centre found
from the galaxy distribution (the cluster models contain galaxies as
well, see van Kampen \& Katgert 1997). Proceeding from number counts to
magnification and convergence estimates as before, we now obtain
estimated convergence profiles. We do this for 32 different backgrounds,
in order to get an estimate for the intrinsic error on the number counts
due to shot noise and background clustering.

This procedure is illustrated in the top panel of Fig.\ \the\Fcountprofex,
where we show the number count profile for one background (thick solid line),
with intrinsic error bars, obtained using all backgrounds, superimposed.
The thin solid line that is also plotted shows the average over 32
different number count profiles obtained by just changing the background
population. The middle panel of Fig.\ \the\Fcountprofex\ shows the
reconstruction of the convergence profile for the single background, along
with the convergence profile obtained directly from the simulation.
The bottom panel shows the same, but now for the averaged profile.
The reconstruction for the latter is obviously better. But despite the
noise, one can clearly get a fair estimate for the convergence profile from
number counts alone.

Finally, we examine how well we can estimate the total mass within a certain
annulus from these count simulations. We simply convert convergence to
surface mass density using the mean redshift of our background galaxies, which
is around 0.8 for the simulations shown, and integrate that to obtain a projected mass within $0.5 h^{-1}$Mpc. We then compare this mass estimate
for the cluster to the mass obtained directly from the numerical cluster model.

Fig.\ \the\Fhistmass\ shows the results of this exercise. The top panel shows
the distribution over estimated masses for the sample of rich clusters that
we considered before. The arrow indicates the mean {\it true}\ mass for this
sample, whereas the vertical lines give the mean for the estimated masses.
The linear mass estimate is a factor of two too large, while all other
estimators give similar values, with the $\gamma=0$ estimator performing
marginally better.
Just one background was used, but we can also consider the effect of changing
the background, i.e.\ cosmic variance, by putting a single cluster in front of
32 different backgrounds. This is shown in the second panel of
Fig.\ \the\Fhistmass, for the third cluster from Table 1.
As this cluster is one of the most massive from the sample, it is no
surprise to see that the linear estimator is now on average a factor of
four too large, with a huge scatter.
The other estimators have a significant scatter as well,
quantifying what was already seen in Fig.\ \the\Fcountmapex. This scatter
decreases for larger convergences, when looking at the same cluster
for larger $\sigma_8$ (shown in the third panel). The cluster is then
more evolved and therefore more massive. The last panel is the same as the
third, but with the linear estimator included, which is now quite far off.
We see that for the more massive clusters the $\gamma\propto\kappa^{1/2}$
estimator works best {\it on average}, but that the scatter among the
estimators is quite similar.

\section{Observed clusters}

\newfig \newcount\Fobs \Fobs=\fignumber
\figps{\fignumber}{D}{\psfig{file=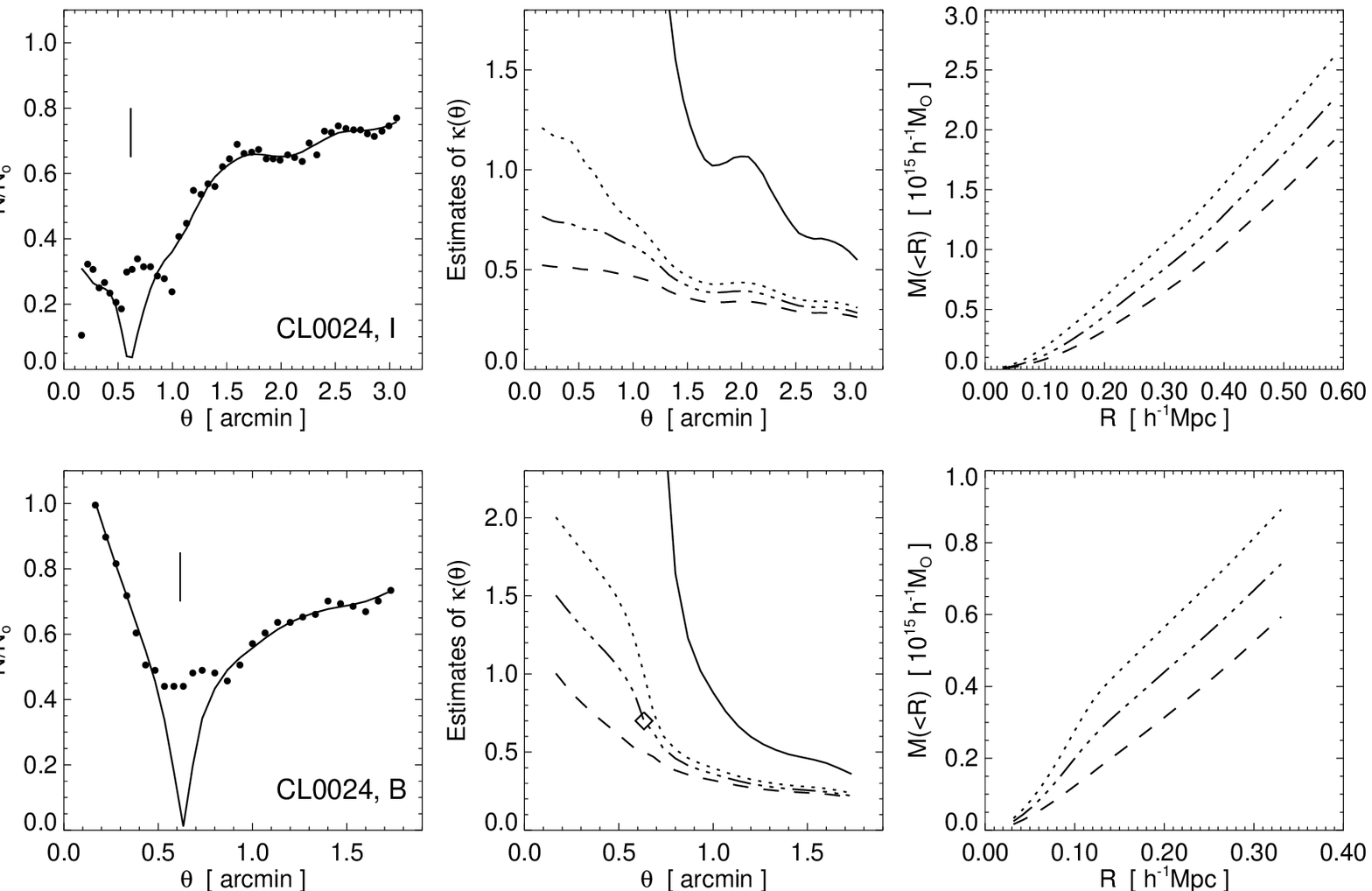,width=17cm,silent=}}
{{\bf Figure \the\fignumber.} Convergence estimates for the clusters
CL0024, in the $I$ and $B$ band.
The left hand panels show the number
count profile, where the dots represent the binned number counts
published by Fort et al.\ (1996) and the solid lines the smoothed
number counts with parity taken into account (see text for details).
The middle panels show the convergence estimates for the various
assumptions made: weak lensing (solid line), $\gamma=0$ (dotted line),
$\gamma=\kappa$ (dashed line), $\gamma\propto\kappa^{1/2}$ (dot-dash
line). The same linetypes have been used to plot the corresponding
cumulative mass profiles in the right hand panel. The position of a
giant arc is indicated by a vertical line. The diamond symbol indicates
a switch in estimators, explained in more detail in the main text.}


\tx We apply the non-linear estimators for the lens convergence to
published observational data, in order to test whether reasonable mass
estimates for real clusters can be made, and how these compare to masses
determined using other, independent techniques. We also like to
establish what the errors are due to the uncertainty in the assumptions
of the non-linear estimators when applied to real data, especially
for the heuristic estimator. After all, so far we only tested the
heuristic estimator on cluster models which were also used to find
that estimator. Presently, data of sufficient quality exists for
two clusters only, CL 0024+1654 and A1689.

\subsection{CL 0024+1654}

\tx Fort, Mellier \& Dantel-Fort (1997) have published number counts
in radial bins for this $z=0.39$ cluster, in both the $I$--band and
the $B$--band. They present counts as surface density profiles, and also
provide the field number counts (which we denote by $N_0$).
For the $I$--band they only count galaxies in the magnitude
range $25<I<26.5$, for the $B$--band in the range $26<B<28$. The
main reason for this is completeness, but also the fact
that they find a field luminosity function with a fairly constant
slope for these intervals: $S_I=0.25\pm0.03$ and $S_B=0.17\pm0.02$
respectively. These values correspond to $\beta_I=-2.67$ and $\beta_B=-1.74$.
Furthermore, a giant arc is observed at 37 arcsec, but its redshift
is unfortunately unknown.
However, Fort et al.\ ({\it ibid.}) argue that it should be close
to the mean for the background population, so we can use its position
to set the parity for our estimators.

The relative number counts $N/N_0$ in $I$ and $B$ are plotted in
Fig.\ \the\Fobs (left panels), along with the various estimators for
the convergence $\kappa$ (middle panels).
Parity was assigned according to the position of the arc, indicated by
a short vertical line in the figure. The number counts were smoothed
with the parities applied, as described in Section 4.2.
This smoothing is really necessary, as the galaxy number
density is fairly small: 6.13 arcmin$^{-2}$ for the $B$--band,
3.32 arcmin$^{-2}$ for the $I$--band, which is even smaller than for the
simulation example shown in Fig.\ \the\Flenscount. The number of bins
used by Fort et al.\ ({\it ibid.}) is too large, given these densities.
The smoothing length that corresponds to these number densities is about
40 arcsec and one arcmin respectively if one would try to contruct a number
count map (see Section 4.2.3). For profiles, the smoothing length
necessary is a decreasing function of radius. However, the number
density of galaxies is increasing with radius (due to the magnification
effect). It is therefore reasonable to use a constant smoothing length
for the profiles. We have adopted 20 arcsec for the $B$--band,
and 30 arcsec for the $I$--band.

The $B$--band data pose a problem for the
$\gamma\propto\kappa^{1/2}$ estimator, as the counts rise all the way back
to $N_0$ at about 10 arcsec. We therefore switch to the `mean' estimator,
described in Section 4.1, at the position where the
$\gamma\propto\kappa^{1/2}$ estimator becomes undefined.
This switch is indicated by a diamond symbol in the bottom
central panel. This procedure is unsatisfactory, of course, but it seems
the best possible alternative for this particular dataset, which is not
of high-quality anyway.

From the convergence profiles we calculate projected mass profiles,
assuming that all background galaxies are at $z_{\rm s}=1$. These are
plotted in the right hand side panels of Fig.\ \the\Fobs, where the
the angular scale is transformed to a physical scale, assuming that
$\Omega_0=1$.
We find that the masses estimated from the $B$-- and $I$--bands are
roughly consistent with each other: for both bands we obtain a total
projected mass within $0.3 h^{-1}$Mpc of approximately
$0.7-0.8\times 10^{15} h^{-1}$M$_\odot$ using the heuristic estimator.
If we treat the $\gamma=0$ and $\gamma=\kappa$ estimators as upper
and lower limits respectively, which in general is not correct, we have
an uncertainty of $0.2\times 10^{15} h^{-1}$M$_\odot$ due to the
uncertainty in the choice of estimator.

Kassiola, Kovner \& Fort (1992) published a mass model for this cluster
that was fitted to various lensing features. They quote a total mass of
about $10^{14} h^{-1}$M$_\odot$ within $0.1 h^{-1}$Mpc.
Our isothermal estimator gives roughly the same mass within that radius,
in both bands. It also gives a mass of $1.6\times 10^{15} h^{-1}$M$_\odot$
within $0.5 h^{-1}$Mpc, which we should consider a lower limit as
the isothermal estimator typically underestimates masses.

Bonnet, Mellier \& Fort (1994) have measured the tangential shear profile
for this cluster, and estimated the mass within $1.5 h^{-1}$Mpc to be
$1-2 \times 10^{15}h^{-1}$M$_\odot$, depending on the assumption for the
density profile. This range contains the mass we find within a radius of
$0.5 h^{-1}$Mpc, so in order to be consistent, we need to assume that
the mass within $1.5 h^{-1}$Mpc is close to the upper limit given by
Bonnet et al.\ ({\it ibid.}), or that most of the cluster mass resides
in the inner $0.5 h^{-1}$Mpc.

However, looking at the counts (left hand panels of Fig.\ \the\Fobs),
we see that the counts never reach the $N_0$ given by Fort et al.\ (1997).
This is not too important for the central regions of the cluster, where the
counts are low anyway, but in the outskirts it yields a significant
contribution to the total mass. If we take a value of $N_0$ to which the
counts do converge, which is 20 per cent smaller than the one taken from
Fort et al.\ ({\it ibid.}), we find that the mass within $0.5 h^{-1}$Mpc is
about $1.0\times 10^{15} h^{-1}$M$_\odot$, while the mass within
$0.1 h^{-1}$Mpc does not change much. This seems a more consistent mass
estimate for CL 0024+1654 from the lens magnification. It is
therefore essential to get a better observational determination for
the field count $N_0$.


Fort et al.\ ({\it ibid.}) also published preliminary data on A370, but
we believe these not to be of sufficient quality to attempt mass estimation,
especially since the position of the minimum in the number counts is far
removed from the position of a giant arc observed in this cluster
(Soucail et al.\ 1988).

%

\subsection{A1689}

\tx We summarize the results for A1689, as published by Taylor et al.\
(1998). The total mass within $0.24 h^{-1}$Mpc was found to be
$0.5\pm0.09\times 10^{15} h^{-1}$M$_\odot$, in fair agreement with X-ray,
virial, and weak shear mass estimates. The number of background galaxies
used was roughly similar to that used in the count simulations presented
in Section 4.2, and the uncertainties are therefore quite similar, i.e.\
on the order of 50 per cent. A1689 has a giant arc, but just like for
CL 0024+1654, the redshift of this arc is unknown, which leaves an
ambiguity in the interpretation.
Note that the total exposure time used to take the A1689 data was quite
short (less than 3 hours in the $V$ and $I$ bands), so an
improvement on this is easily achieved.

\section{Conclusions and discussion}

\tx In order to be able to estimate the lens convergence $\kappa$ from
the lens magnification $\mu$, we searched for a realistic
approximation between either of these quantities and the lens shear
$\gamma$, which needs to be taken into account in the strong lensing
regime. We looked at simple analytical lens models with exact
relations between the three lens quantities, notably the $\gamma=0$
approximation, corresponding to a uniform sheet of matter, and
the isothermal model, which has $\kappa=\gamma$.

As these do not provide adequate descriptions
for neither observed nor simulated clusters,
we have studied the lensing properties of a complete catalogue
of galaxy cluster models to find an approximation for the shear
which still allows a simple inversion from magnification to
convergence. A approximation that does just that is
$\gamma\propto\kappa^{1/2}$, which leads to a simple two-caustic
magnification-convergence relation. A disadvantage of this
approximation is that two parities have to be set, which can be
a problem for observational data, as the sign of the magnification
cannot be directly measured.

We also discussed the iterative
technique, where one starts with any of the estimators mentioned
to obtain an estimate for the shear using the thin lens approximation,
and then uses that shear to calculate a new convergence estimate.
However, this method is not likely to converge for observational
data because the dominant quadrupole structure of the shear field
will not be present in the observed magnification field, which
usually needs to be smoothed fairly heavily. The iterative estimate
is still useful, however, if one performs a few steps only, or preferrably
just one.

Two types of tests were performed on all estimators considered:
the ideal case for which full knowledge of the magnification
distribution exists (including its sign),
and the case where we start from mock lensed background galaxy
counts, mimicking intrinsic problems like the clustering of these
galaxies and shot noise due their limited number.

The first type of tests have shown that if the magnification is
perfectly known, the mass distribution of these cluster models can
be fairly well reconstructed using the estimator based on a the
relation and convergence of the form $\gamma\propto\sqrt\kappa$,
or the iterative estimator operated for one step only.
The isothermal estimate generally underestimates the surface
mass density for $\kappa>0.3$, and overestimates for smaller $\kappa$.
The $\gamma=0$ estimator overestimates for all $\kappa$ up to
the first caustic, and can therefore be considered a strict lower
limit for these $\kappa$.

The second type of tests aimed at mimicking observational data, i.e.\
going from number counts to convergence estimates. We showed that it
is still possible to reconstruct the convergence for simulated clusters,
although the intrinsic uncertainties become significant.
We have also demonstrated that the mass estimated using the weak lensing
approximation is at least twice the real mass, thus showing that one
really needs to go beyond the weak lensing approximation to get
sensible cluster mass estimates.

We used published number counts for the cluster CL 0024+1654 (Fort et al.\
1997) to illustrate the non-linear mass estimation technique. We showed
that the total mass estimated compares fairly well to estimates from other
techniques, even though the quality of the data is relatively poor.
The uncertainties in the mass found for both this cluster and
for A1689 (Taylor et al.\ 1997) are quite large still, but there are
many ways to reduce these. We can get
(photometric) redshifts for the background galaxies, find a better
value for the field number count, obtain the redshifts of the arcs,
and use wavelengths for which the luminosity function is either much steeper
of much shallower that the slope of 0.4 for which there is no variation in
number counts at all. Also, we should go to fainter magnitudes, in order to
get higher background galaxy number densities, which minimizes the clustering
and shot noise problems. Furthermore, the lens becomes stronger as the
convergence gets larger due to the higher mean background galaxy redshift.

In concluding, we showed that mass estimation from lens magnification
in the strong lensing regime is possible. We seem to find
consistent mass estimates for both CL 0024+1654 and A1689 with only limited
observational data available. This should give us encouragement to obtain
better data for these and other clusters. Obviously, in combination with
other mass estimates the lens magnification technique will be even more promising.

\section*{acknowledgments}

\tx The author thanks Andy Taylor for supplying the mock background
galaxy distributions and for useful discussion and comments,
John Peacock, Alan Heavens, Jens Hjorth, Cedric Lacey and Frank Pijpers for
useful discussion and comments, and acknowledges an European Community
HCM Research Fellowship for financial support.
This work was supported in part by Danmarks Grundforskningfond through
its funding of the Theoretical Astrophysics Center.

\section*{references}

\bibitem Bartelmann M., Weiss A., 1994, A\&A, 287, 1
\bibitem Bartelmann M., 1995, A\&A, 303, 643
\bibitem Bartelmann M., Steinmetz M., Weiss A., 1995, A\&A, 297, 1
\bibitem Bartelmann M., Narayan R., Seitz S., Schneider P., 1996, ApJ, 464, L115
\bibitem Broadhurst T.J., Taylor A.N., Peacock J.A., 1995, ApJ, 438, 49
\bibitem Bonnet H., Mellier Y., Fort B., 1994, ApJ, 427, L83
\bibitem Coles P., Jones B.J.T., 1991, MNRAS, 248, 1
\bibitem Fort B., Mellier Y., 1994, A\&AR, 5, 239
\bibitem Fort B., Mellier Y., Dantel-Fort M., 1997, A\&A, 321, 353 
\bibitem Gorenstein M.V., Falco E.E., Shapiro I.I., 1988, ApJ, 327, 693
\bibitem van Kampen E., 1994, PhD Thesis, Rijksuniversiteit Leiden
\bibitem van Kampen E., 1996, in Mart\`\i nez, V.J., Coles, P., eds., Mapping,
Measuring, and Modelling the Universe, ASP Conference Series Vol. 94, p. 265
\bibitem van Kampen E., 1997, in Clarke D.A., West M.J., eds., Computational
Astrophysics, ASP Conference Series Vol. 123, p. 231
\bibitem van Kampen E., Katgert P., 1997, MNRAS, 289, 327
\bibitem van Kampen E., 1998, in preparation
\bibitem Kaiser N., 1995, ApJ, 493, L1
\bibitem Kaiser N., 1996, in Lahav O., Terlevich E., Terlevich R.J., eds.,
'Gravitational Dynamics', Proc. of the 36th Herstmonceux Conference.
Cambridge Univ. Press, Cambridge, p. 181
\bibitem Kaiser N., Squires G., 1993, ApJ, 404, 441
\bibitem Kassiola A., Kovner I., Fort B., 1992, ApJ, 400, 41
\bibitem Katgert P., Mazure A., Jones B., den Hartog R., Biviano A.,
Dubath P., Escalera E., Focardi P., Gerbal D., Giuricin G.,
Le F\`evre O., Moles O., Perea J., Rhee G., 1995, A\&A, 310, 8
\bibitem Kochanek C.S., Blandford R.D., 1991, ApJ, 375, 492 
\bibitem Mazure A., Katgert P., den Hartog R., Biviano A., Dubath P.,
Escalera E., Focardi P., Gerbal D., Giuricin G., Jones B.,
Le F\`evre O., Moles O., Perea J., Rhee G., 1995, A\&A, 310, 31
\bibitem Metcalfe N., Shanks T., Fong R., Roche N., 1995, MNRAS, 273, 257
\bibitem Schneider P., 1995, A\&A, 302, 639
\bibitem Schneider P., Ehlers J., Falco E.E., 1992, Gravitational Lenses,
Berlin: Springer
\bibitem Schneider P., Seitz C., 1995, A\&A, 297, 411
\bibitem Silverman B., 1986, Density Estimation for Statistics and Data
Analysis, London: Chapman and Hall
\bibitem Soucail G., Mellier Y., Fort B., Mathez G., Cailloux M., 1988,
A\&A, 191, L19
\bibitem Taylor A.N., Dye S., 1998, submitted to MNRAS
\bibitem Taylor A.N., Dye S., Broadhurst T.J., Ben\'\i tez N., van Kampen E.,
1998, ApJ, in press
\bibitem Wilson G., Cole S., Frenk C.S., 1996, MNRAS, 280, 199
\section*{Appendix A: Proof that spherical lensing potentials cannot
have $\gamma\propto\kappa^{1/2}$}

\eqnumber=1

\tx For spherically symmetric lensing potentials, both the lens
convergence $\kappa(x)$ and magnification $\mu(x)$ are functions of the
deflection angle ${\bf\alpha}({\bf x})$ and its first derivative
(eg.\ Schneider et al.\ 1992):  $$\kappa(x)={1\over
2}\Bigl({\alpha(x)\over x}+
  {{\rm d}\alpha(x)\over{\rm d}x}\Bigr) \ , \eqno(A\neweq)$$ and
$$\eqalign{\mu^{-1}(x) =& \Bigl(1-{\alpha(x)\over x}\Bigr)
	\Bigl(1-{{\rm d}\alpha(x)\over{\rm d}x}\Bigr) \cr
	=&\Bigl({1\over c}-{\alpha(x)\over cx}\Bigr) \Bigl(c-c{{\rm
	d}\alpha(x)\over{\rm d}x}\Bigr)\ . \cr} \eqno(A\neweq)$$ So if
we want to have $\mu^{-1}=(c^{-1}-\kappa)(c-\kappa)$ (e.g.\ eq.\ 
\the\muinv), we need to have the following two equalities:
$${\alpha(x)\over cx} = c{{\rm d}\alpha(x)\over{\rm d}x} =
   {1\over 2}\Bigl({\alpha(x)\over x}+ {{\rm d}\alpha(x)\over{\rm
  d}x}\Bigr)\ . \eqno(A\neweq)$$
Eliminating ${\rm d}\alpha(x)/{\rm d}x$, we have the following
condition for the approximation $\gamma\propto\kappa^{1/2}$ to hold:
$$(c^2-2c+1) \alpha(x) = 0 \ .  \eqno(A\neweq) $$
Besides the solution $\alpha(x) = 0$, corresponding to a sheet of
matter, this condition is not met for any real $c$.

%
%

\end